\documentclass[aps,reprint,twocolumn,superscriptaddress,english,longbibliography,floatfix]{revtex4-2}
\usepackage{amssymb}
\usepackage{graphicx}
\usepackage{amsmath}
\usepackage{epsfig}
\usepackage{array}
\usepackage{multirow}
\usepackage{color}
\usepackage{esint}
\usepackage{bm}
\usepackage{bbm}
\usepackage{epstopdf}
\usepackage{mathtools}
\usepackage{soul}
\usepackage[dvipsnames]{xcolor}
\usepackage{tikz}
\usepackage[bookmarks=true,pdftex,colorlinks=true,urlcolor=blue,linkcolor=black,citecolor=blue,breaklinks=true,hypertexnames=false]{hyperref}
\usepackage{multibib}
\usepackage{physics}
\usepackage{float}
\begin{document}

	\setcounter{equation}{0} \setcounter{figure}{0}
	\setcounter{table}{0} \setcounter{page}{1} \makeatletter
    \title{Quantum Quenches across the Bose-glass Transition}
	
	\author{Bo Song}\email{bsong@pku.edu.cn}
	\affiliation{Cavendish Laboratory, University of Cambridge, J.J. Thomson Avenue, Cambridge CB3 0HE, United Kingdom\looseness=-1}
	\affiliation{State Key Laboratory for Mesoscopic Physics and Frontiers Science Center for Nano-optoelectronics, School of Physics, Peking University, Beijing 100871, China}
    
	\author{Shaurya Bhave}
	\affiliation{Cavendish Laboratory, University of Cambridge, J.J. Thomson Avenue, Cambridge CB3 0HE, United Kingdom\looseness=-1}
    \affiliation{Department of Physics, University of Strathclyde, Glasgow G4 0NG, United Kingdom\looseness=-1}
   
    \author{Emmanuel Gottlob}
	\affiliation{Cavendish Laboratory, University of Cambridge, J.J. Thomson Avenue, Cambridge CB3 0HE, United Kingdom\looseness=-1}
    
	\author{Leanne Reeve}\email{lcr37@cam.ac.uk}
	\affiliation{Cavendish Laboratory, University of Cambridge, J.J. Thomson Avenue, Cambridge CB3 0HE, United Kingdom\looseness=-1}

	\author{Ulrich Schneider}\email{uws20@cam.ac.uk}
	\affiliation{Cavendish Laboratory, University of Cambridge, J.J. Thomson Avenue, Cambridge CB3 0HE, United Kingdom\looseness=-1}

	\date{\today}

    \begin{abstract}
        Due to their intrinsic interplay between long-range order and quasi-disorder, quasicrystalline systems provide a rich platform for investigating novel quantum phenomena. Here we study the non-equilibrium dynamics following quantum quenches across the superfluid to Bose-glass transition using ultracold bosons in an optical quasicrystal. Fast quenches into the Bose glass regime induce a quantum walk in momentum space that spreads over increasingly higher momentum orders. Conversely, quenches into the superfluid regime initiate a real-space quantum walk of initially localized atoms, forming a light-cone-like structure bounded by Lieb–Robinson limits. Characteristic timescales reveal strong links to the underlying Hamiltonian and are governed primarily by quasi-disorder strength in the Bose glass phase, and by tunneling strength in the superfluid phase. Finally, we analyze slower quenches into deeper lattices and observe the decay of coherence across the phase transition. 
    \end{abstract}
    
    \maketitle

\section{Introduction}
The nonequilibrium dynamics of quantum many-body systems are central to many areas of modern physics. Prominent examples include quantum thermalization~\cite{rigol2008thermalization,eisert2015quantum}, quantum chemistry~\cite{kassal_polynomial-time_2008,lu_simulation_2011,arguello-luengo_engineering_2021}, the dynamics of dynamical gauge fields~\cite{banerjee_atomic_2012,marcos_two-dimensional_2014,barbiero_coupling_2019}, and the dynamics of phase transitions~\cite{bernien2017probing,zhang2017observation,song_realizing_2022}. Ultracold atoms provide an ideal platform for studying these dynamics, being both highly tuneable and exhibiting coherent dynamics over long and experimentally accessible timescales~\cite{daley_practical_2022}. Current examples include many-body localization~\cite{schreiber_observation_2015, choi2016exploring,abanin2019colloquium}, time crystals~\cite{zhang2017observationTimeCrystal,choi2017observation}, turbulence~\cite{navon_emergence_2016}, and coarsening dynamics~\cite{martirosyan_universal_2025,manovitz_quantum_2025}, amongst others \cite{langen_ultracold_2015,schafer_tools_2020}. A natural way to induce far-from-equilibrium dynamics is via quantum quenches, i.e., fast or sudden changes of the Hamiltonian that take the system far out of equilibrium and thereby induce complex dynamics. 

Disorder often plays an essential role in nonequilibrium phenomena, from modifying transport properties to stabilizing the electrical properties of the quantum Hall effect~\cite{levi2012hypertransport,joynt_conditions_1984,pruisken_universal_1988,seye_localization_2025}. One famous example is Anderson localization, in which sufficiently strong disorder inhibits wave propagation and thus leads to localization and vanishing conductivity~\cite{anderson1958absence,abrahams201050}. In bosonic systems, introducing interactions in disordered systems can give rise to a novel localized state, the Bose glass (BG), which is an insulating but compressible phase without long-range order. While originally introduced as a ground state~\cite{giamarchi1988anderson,fisher1989boson}, the Bose glass was later found to extend to finite energy densities~\cite{ciardi_finite-temperature_2022,michal_finite-temperature_2016,bertoli_finite-temperature_2018,zhu_thermodynamic_2023}.  
A rich alternative to disordered systems are quasicrystals, systems that are long-range ordered, yet non-periodic. In optical quasicrystals, i.e., quasiperiodic optical lattices for ultracold atoms \cite{viebahn2019matter}, the lack of translational symmetry leads to site-dependent on-site energies and tunneling coefficients~\cite{gottlobHubbardModelsQuasicrystalline2023}. 

Localization phenomena have been studied with ultracold atoms in both randomly disordered and quasiperiodic systems, including Anderson localization~\cite{billy2008direct,roati2008anderson,kondov2011three,jendrzejewski2012three}, Bose glasses in one-dimensional (1D)~\cite{fallani2007ultracold,d2014observation, gadway2011glassy} and 3D disordered systems~\cite{pasienski2010disordered, meldgin2016probing}, and, more recently, in 2D quasiperiodic~\cite{BG2022} and disordered~\cite{koehn2026} systems. 

In this work, we experimentally study the nonequilibrium dynamics following quantum quenches across the phase transition between the superfluid and the recently realized Bose glass phase in a 2D optical quasicrystal~\cite{BG2022}. Using time-of-flight (TOF) imaging, we study the momentum distribution of the many-body state, which provides access to the coherence dynamics. We observe two distinct dynamical behaviours depending on the direction of the quench: When the system is quenched into the strongly-disordered Bose glass regime, atoms follow a quantum walk in momentum space spreading to successively higher diffraction orders~\cite{viebahn2019matter} and causing oscillations of the zero-momentum peak with a frequency controlled by the spread in on-site energies. In contrast, when the final regime is a low-disorder superfluid, we observe slower dynamics dominated by the real-space tunneling of particles. We further investigate the dynamics of the system following a slow quench from a superfluid to different final lattice depths and discuss how entropy transport within the inhomogeneous system controls the coherence decay at the center of the cloud.

\subsection{Quasiperiodic Bose-Hubbard model and weakly-interacting phases}
The two-dimensional (2D) optical quasicrystal is formed by overlapping four (periodic) 1D lattices at 45$^\circ$ to one another in the horizontal ($x,y$) plane~\cite{viebahn2019matter,sbroscia2020observing}. These lattices are generated using blue-detuned light at a wavelength of $\lambda_{\text{latt}}\approx725$\,nm, and throughout this paper we will refer to the corresponding wavenumber and single-photon recoil energy as ${k_{\text{latt}}=2\pi/\lambda_{\text{latt}}}$ and ${E_r=\hbar^2k_{\text{latt}}^2/2m}$ respectively. Here $m$ denotes the atomic mass of the used bosonic potassium ($^{39}$K). This quasicrystalline horizontal lattice gives rise to a potential with an eight-fold rotational symmetry that hence cannot be periodic. It is combined with a sufficiently deep lattice along the vertical ($z$) direction (depth $V_z=20E_r$) to split the system into essentially independent 2D layers. In the tight-binding approximation, the Hamiltonian takes the form of a quasiperiodic Bose-Hubbard model~\cite{gottlobHubbardModelsQuasicrystalline2023},

\begin{equation} \label{eq:Hubbard}
    \hat{H}=\sum_{i,j}-J_{i,j}\hat{a}_i^\dag\hat{a}_j+\sum_{i}\frac{U_i}{2}\hat{n}_i(\hat{n}_i-1) + \sum_{i}(\epsilon_i+V^{\text{trap}}_{i})\hat{n}_i,
\end{equation}
where $\hat{a}_i^\dag$ ($\hat{a}_i$) creates (annihilates) a particle on site $i$, and $\hat{n}_i = \hat{a}_i^\dag \hat{a}_i$ is the corresponding number operator. $J_{i,j}$ denotes the tunneling element between sites $i$ and $j$, $U_i$ is the on-site interaction at site $i$, $\epsilon_i$ the on-site energy due to the quasiperiodic lattice, and $V^{\text{trap}}_i$ gives the contribution to the on-site energy from an underlying harmonic trap used to confine the atoms. Note that $J_{i,j},\,U_i$ and $\epsilon_i$ all follow quasiperiodic patterns. The quasi-disorder strength $\Delta=\text{max}(\epsilon_i)-\text{min}(\epsilon_i)$ is defined as the range of the on-site energies $\epsilon_i$ and, for $V_\text{latt}\gtrsim2\,E_r$, increases approximately linearly with lattice depth~\cite{gottlobHubbardModelsQuasicrystalline2023}.

\begin{figure}[ht]
    \centering
    \includegraphics[width=0.48\textwidth]{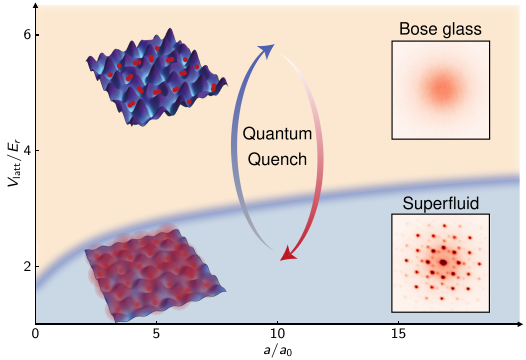}
    \caption{\textbf{Schematic of quantum quenches between superfluid  and Bose glass in an optical quasicrystal.} When the quasicrystal is shallow, the ground state of weakly interacting bosonic atoms is a superfluid (SF, bright blue) and the atoms are coherently delocalized over the whole lattice, whereas for a deep lattice, atoms are localized due to quasi-disordered on-site energy terms and the system enters a Bose glass (BG, yellow) regime. Right insets: a SF has sharp interference peaks in time-of-flight images, signalling long-range coherence, whereas a BG has a broad momentum distribution without any sharp features. By rapidly ramping the lattice potential of the quasicrystal, we quench between these two regimes and monitor the resulting time evolution of the matter-wave field. 
    \label{fig:Quench_Schematic}}
\end{figure}

In the shallow lattice regime with small interactions tunneling dominates and the ground state is superfluid, see Fig.~\ref{fig:Quench_Schematic}. For deep quasicrystalline lattices, the variation in the quasi-disordered on-site energy becomes the largest energy scale and the Hamiltonian is dominated by $\hat{H}= \sum_{i}\epsilon_i\hat{n}_i$. As a consequence, the weakly-interacting ground state is the Bose glass (yellow regime in Fig.~\ref{fig:Quench_Schematic}) and the dynamics are dominated by the quasi-disorder~\cite{yang2017dynamical,gottlobHubbardModelsQuasicrystalline2023}. 
For periodic systems, in contrast, the Hamiltonian in the deep lattice limit can be approximated to $\hat{H}= U_i\hat{n}_i(\hat{n}_i-1)/2$, leading to Mott insulating ground states. After a quench from a shallow lattice superfluid into this regime (SF$\rightarrow$MI), the matter-wave field oscillates at a frequency determined by the interaction strength~\cite{greiner2002collapse,will2010time}. The evolution after the opposite quench into a weak lattice (MI or BG$\rightarrow$SF) is instead dominated by the tunneling of atoms between adjacent sites.

Taking into the account the full quasiperiodic Hamiltonian, the ground state will change from a SF into a BG at a critical lattice depth that for weakly-interacting bosons increases with density, since weak repulsive interactions can effectively screen the system's quasi-disorder, inducing tunneling resonances and delocalization. In the presence of an underlying harmonic trap, the higher density part in the centre of the atomic cloud will hence localize at a deeper critical lattice depth than the lower density edges, as demonstrated in~\cite{BG2022}. Throughout this paper, we will refer to the part of the phase diagram where the whole system is localized as the Bose glass regime, and that where at least some region of the cloud is still delocalized as the superfluid regime, see Fig.~\ref{fig:Quench_Schematic}.  

\subsection{Experimental sequence}
The experiment starts with the preparation of a BEC of around $10^5$ bosonic $^{39}$K atoms in the $|F=1,m_F=1\rangle$ state. These atoms are loaded into the quasicrystal as in~\cite{BG2022} by ramping up the lattice depth exponentially in $45\,$ms with a time constant of $10\,$ms. The scattering length is set to approximately $10a_0$, where $a_0$ is the Bohr radius, using the broad Feshbach resonance at $402.74\,$G~\cite{etrych_pinpointing_2023} - this scattering length is used for all experiments described in this paper unless otherwise specified. After holding the atoms in this initial lattice for $10\,$ms, we rapidly ramp (quench) the depth $V_{\text{latt}}$ of the horizontal lattices, using either a \textit{fast}, almost instantaneous $10\,$\textmu s quench or a \textit{slow} $290\,$\textmu s quench. We then hold the lattice depth at this final value for a varying time and study the resulting dynamics. 
For fast quenches to the SF regime, or slow quenches inducing long-term dynamics, we finally implement an additional short booster stage~\cite{BG2022} in a deeper lattice. This stage is short enough not to change the coherence properties of the system but provides a tighter on-site confinement and thereby enhances the brightness of high-order diffraction peaks, which improves the extraction of the coherence of the system. See appendix~\ref{SuppMat:sequence} for further details of the sequences used. In all cases, we measure the final momentum distribution of the atoms using absorption imaging after a time-of-flight (TOF) of $9\,$ms, extracting the optical density (OD) as a measure of the column density of atoms integrated along the $z$ direction.

\section{Fast quenches}
We first study the evolution following a fast quench, where the lattice depth is ramped across the localization transition within $10\,$\textmu s, which is fast compared to the tunneling rate. For $V_{\text{latt}}^{\text{final}}>1\,E_r$,  where the lowest band is separated by a gap~\cite{gottlobHubbardModelsQuasicrystalline2023}, this ramp is at the same time slow enough to avoid significantly populating the excited bands.
\begin{figure*}[ht]
    \centering
    \includegraphics[width=1.0\textwidth]{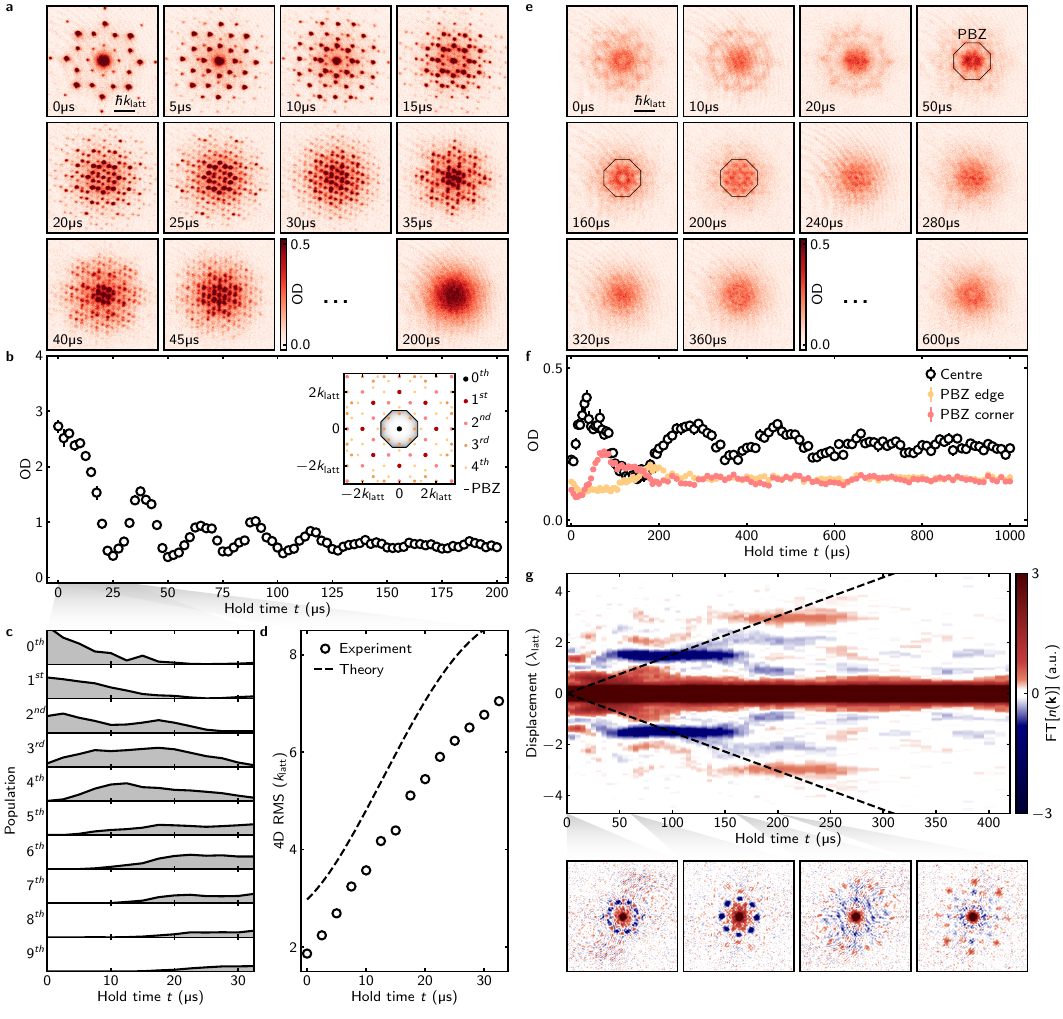}
    \caption{\textbf{Time evolution of matter waves after quantum quenches to SF and BG regimes.} (a) TOF images (averaged over 5 repetitions) after a quench from the SF regime ($V_{\text{latt}}=2E_r$) to the BG regime ($V_{\text{latt}}=7E_r$), showing dynamics involving reciprocal lattice vectors of different orders. (b) Optical density (OD) at $\mathbf{k}=\mathbf{0}$ versus hold time, showing decaying oscillations. Error-bars show the standard-error in the mean at each time-step, and are smaller than the markers when not visible. Inset shows the positions of the momentum peaks up to $4^{th}$ order, and the boundary of the first pseudo-Brillouin zone (PBZ). (c) Evolution of the population of momentum peaks up to $9^{th}$ order, showing progressive population of higher-order peaks. See appendix~\ref{SuppMat:high_order_peak_extraction} for details of the fitting procedure used to extract these populations. (d) RMS momentum of the density distribution in the equivalent four-dimensional hypercubic lattice (see~\cite{viebahn2019matter}). This analysis reveals a ballistic quantum walk in the 4D hypercubic reciprocal lattice, which we find agrees well with numerical simulations based on a tight-binding model of the quasicrystal up to a time offset of roughly $5\,$\textmu s. (e) TOF images (averaged over 5 repetitions) after a quench from the BG regime ($V_{\text{latt}}=6E_r$) to the SF regime ($V_{\text{latt}}=1E_r$). The total atom numbers in these images are lower than those seen in (a), which we attribute to increased atom loss rates during the loading process and initial lattice hold when loading into deeper lattices. (f) OD at $\mathbf{k}=\mathbf{0}$ as well at PBZ corners and edges versus time, exhibiting much slower frequencies than the quenches to the Bose glass regime. Error-bars show the standard-error in the mean at each time-step, and are smaller than the markers when not visible. (g) Trap-averaged real-space correlation function, obtained by Fourier transforming the momentum space density distribution. Plots show individual images of the 2D correlations (bottom) and cuts through the origin (averaged across the four lattice directions) versus time (top). The correlations exhibit a light-cone-like ballistic expansion, in good agreement with the analogue of the maximum group velocity (dashed lines). \label{fig:Quench_SF_BG}}
\end{figure*}

\subsection{Fast quenches into Bose glass regime}
In the quench from the SF to the BG regime, the system starts in a SF state with long-range phase coherence, giving rise to discrete sharp peaks in momentum space. However, after the quench into the deep BG regime, the quasiperiodic on-site energies dominate over tunneling and the atoms lose their original coherence and become localized, see Fig.~\ref{fig:Quench_SF_BG}(a). In contrast to most other systems, the localization does not proceed via a broadening of the momentum peaks, but rather by the proliferation of more and more sharp momentum peaks. 

This can be understood as a quantum walk in momentum space where the long-range ordered nature of the quasicrystal ensures that only discrete momentum classes are coupled~\cite{viebahn2019matter}. The inset in Fig.~\ref{fig:Quench_SF_BG}(b) illustrates this: the eight-fold quasicrystal couples momentum zero (0th order) to eight 1st-order momenta that are in turn coupled to more 2nd-order momenta and so on. In contrast to periodic lattices, higher momentum orders do not necessarily correspond to higher kinetic energies, but rather give rise to a dense set of discrete momenta of successively higher orders~\citep{viebahn2019matter}. As a consequence, there is no upper bound on the order of momentum peaks that atoms can reach and the dynamics is dominated by a flow of population to successively higher-order momentum peaks, see  Fig.~\ref{fig:Quench_SF_BG}(c). At long times (200$\,\mu$s), the peaks become too dense to be resolved and the observed distribution becomes featureless and broad, characteristic of localized states. In a real-space picture, the same dynamics can be understood in terms of the quasi-disordered on-site energy, which in the absence of tunneling gives rise to differential phase evolutions for atoms on different sites leading to a loss of spatial coherence. 

In order to extract the characteristic timescales of these dynamics, we focus on the oscillating occupation of the zero-momentum peak $n(\mathbf{k}=\mathbf{0},t)$ shown in Fig.~\ref{fig:Quench_SF_BG}(b) and extract the dominant frequency using fits to its Fourier transform, see appendix~\ref{SuppMat:FreqExtraction}. Repeating this measurement for several final lattice depths and plotting the results in Fig.~\ref{fig:Quench_SF_BG_freq}, we observe that the dominant frequency increases approximately linearly with lattice depth in broad agreement with the quasi-disorder strength $\Delta$. We also numerically reproduce the quantum quench with a mean-field simulation (see appendix~\ref{SuppMat:theory} for details) and find agreement in the dominant frequency (grey band in Fig.\ref{fig:Quench_SF_BG_freq}). 
For these short times, the dynamics are not sensitive to the interaction strength and initial lattice depth, as the quasi-disorder is the dominating energy scale at the final lattice depth, see insets. This is further confirmed by comparing the dynamics at different scattering lengths and initial lattice depths, see appendix~\ref{SuppMat:InitialState}.

\subsection{Fast quenches into superfluid regime}
In the complementary quench from the BG to the SF regime, shown in Fig.~\ref{fig:Quench_SF_BG}(e),  the initial state is a strongly-localized Bose glass with a broad momentum distribution. The structure visible for $t=0\,\mu s$ reflects dynamics during the quench ramp. Once tunneling is strong, the atoms undergo a transient buildup of short-range coherence as they spread throughout their neighbouring sites, generating patterns in their momentum distribution. This coherence is however eventually lost as the now ergodic system thermalizes to a temperature above the superfluid transition temperature and the momentum patterns disappear. 

\begin{figure*}[t]
    \centering
    \includegraphics[width=1.0\textwidth]{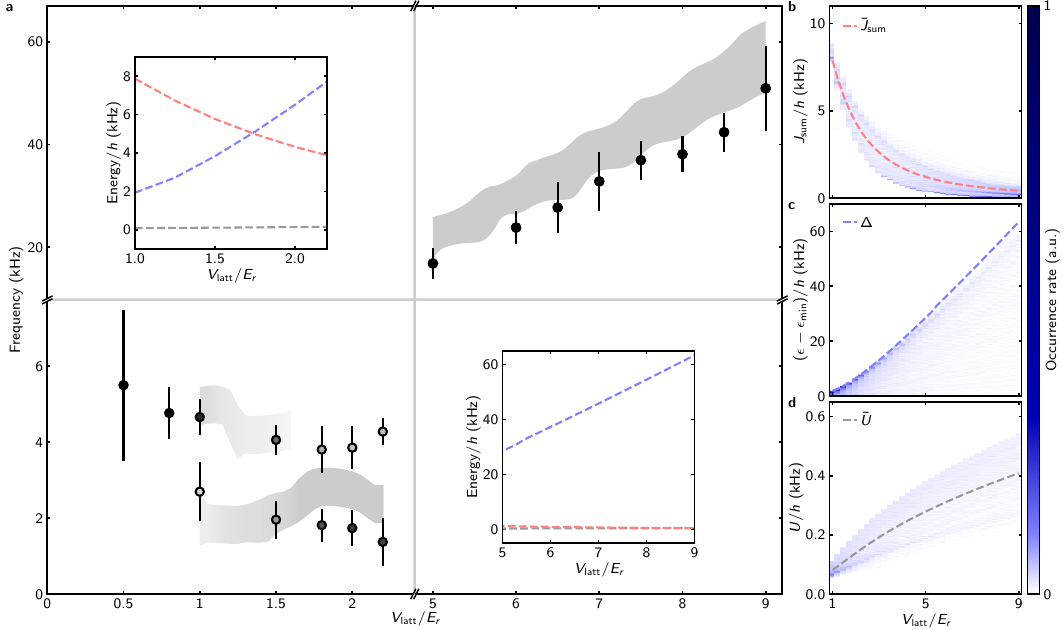}
    \caption{\textbf{Dominant frequencies following fast quenches between the SF and BG regime.} In (a) we plot the dominant frequencies following quenches to different final lattice depths. Error bars represent the standard deviation of the Gaussian which the peak was fitted with. Where two peaks are present, we indicate the stronger peak with a darker color. We employ the same fitting methods to extract the peak locations from a mean-field simulation (which is detailed in appendix~\ref{SuppMat:theory}), and plot as a grey band the region bounded by the extracted frequency $\pm$ one standard deviation of the fitted Gaussian. Again, a darker grey indicates a higher power relative to the other peak in cases where there are two peaks. For the quench from a SF ($V_{\text{latt}}=2E_r$) to the BG regime, the frequency of the oscillation increases with lattice depth, which we attribute to the increase in the quasi-disorder of on-site energies. For quenches from the BG ($V_{\text{latt}}=6E_r$) to the SF regime, there can be multiple frequencies involved in the dynamics, both on the order of the tunneling amplitudes. As the final lattice depth is increased, the dominant frequency decreases, and the relative strengths of the higher and lower frequency peaks are reversed (see appendix~\ref{SuppMat:FreqExtraction}). (b), (c) and (d) show the distribution of the quasiperiodic Hubbard parameters for varying lattice depth in blue, where $J_{\text{sum},\,i}=\sum_j |J_{ij}|$, and $\epsilon_{\text{min}}$ corresponds to the lowest on-site energy in the lattice. Dashed lines show the mean of these values across all lattice sites, and are also plotted in the insets of (a). In the BG regime the quasi-disorder in the on-site energies (blue) dominates, while tunneling (red) dominates for weak lattices, and both are comparable for lattice depths of $1.5-2\,E_r$.
    \label{fig:Quench_SF_BG_freq}}
\end{figure*}

In contrast to the previous quench, these momentum dynamics cannot be understood by considering low-order momenta. Peaks instead appear at different momenta such as the corners of the first pseudo-Brillouin zone (PBZ)~\cite{man_experimental_2005, spurrier_semiclassical_2018} that do not correspond to low-order momentum peaks. We analyse these structures using the spatial Fourier transform of the momentum density $\text{FT}[n(\mathbf{k})]$ which as shown in appendix~\ref{SuppMat:real_space_autocorr} gives the trap-averaged real-space correlation function: 
\[g(\mathbf{r})=\int\left\langle\hat{\psi}^{\dag}(\mathbf{r}')\hat{\psi}(\mathbf{r}'+\mathbf{r})\right\rangle d\mathbf{r}',\]
where $\hat{\psi}(\mathbf{r})$ is the field annihilation operator at position $\mathbf{r}$. This observable is shown in Fig.~\ref{fig:Quench_SF_BG}(g), and provides access to the inter-site correlations (coherences) $\langle\hat{a}_i^\dagger \hat{a}_j\rangle $ averaged across the whole system, see appendix~\ref{SuppMat:real_space_autocorr}. Fig.~\ref{fig:Quench_SF_BG}(g) clearly shows that the observed dynamics correspond to a light-cone-like spreading of real-space correlations, compatible with the ballistic expansion of the initially localized particles (dashed lines)~\cite{Cheneau2012}. The speed of the spreading is in good agreement with the analogue of the maximum group velocity calculated using the tunneling rates and intersite distances of the quasicrystal (see appendix~\ref{SuppMat:real_space_autocorr}), indicating that this dynamics saturates the Lieb-Robinson bound~\cite{lieb_finite_1972}. Hence we find that, in contrast to the quantum walk in momentum space observed in the quench from the SF to the BG regime, here we initially observe a quantum walk in real space.

We study the characteristic timescale of this real-space quantum walk by again Fourier transforming  $n(\mathbf{k}=\mathbf{0},t)$ (e.g.\ Fig.~\ref{fig:Quench_SF_BG}(f)) in time. Depending on the final lattice depth, we observe either one or two peaks in the spectrum and extract their locations using gaussian fits, see appendix~\ref{SuppMat:FreqExtraction}. As shown in Fig.~\ref{fig:Quench_SF_BG_freq}, the dynamics are markedly slower than those following the quench to the BG regime. This reflects the fact that the dynamics is due to tunneling, which is a far smaller energy scale than the quasi-disorder after the quench to the BG regime. As expected, the dominant frequency decreases with final lattice depth in line with the decrease in tunneling amplitudes and in quantitative agreement with mean-field simulations.

\begin{figure*}[t]
    \centering
    \includegraphics[width=1\textwidth]{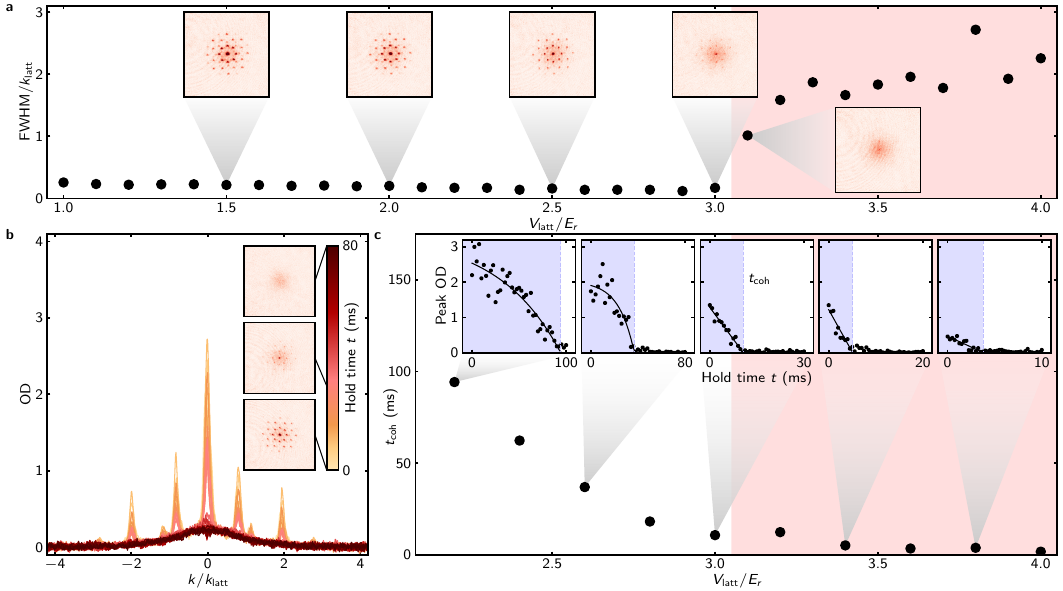}
    \caption{
    \textbf{Coherence time after a slow quench reflects the BG phase transition.} (a) Coherence after adiabatically loading atoms into different lattice depths without any quench at a scattering length of $10a_0$. The full-width-at-half-maximum of the central peak (as measured in~\cite{BG2022}) is plotted as a function of lattice depth, showing that the phase transition occurs at around $V_{\text{latt}}=3-3.1E_r$ at the centre of the cloud, with the shaded area corresponding to the regime where the cloud is fully localized. Inset: TOF images for different lattice depths. (b) Dynamics of the system after a slow quench from a SF ($V_{\text{latt}} = 2 E_r$) to a lattice depth of $2.6E_r$ in $290\,$\textmu s. The main plot shows cuts through the centre of the TOF images averaged across a window of width $0.1k_0$ for various hold times after the slow quench. The sharp diffraction peaks gradually decay until only a broad background remains for long enough hold times. TOF images for various hold times are shown as insets. (c) Repeating the slow quench for varying lattice depths, we in each case measure the time $t_{\text{coh}}$ up to which some degree of coherence persists. To this end, we fit the central cuts from (b) with a bimodal fit and extract the height of the sharp central peak (Peak OD) and plot its evolution in the insets. The times  $t_{\text{coh}}$, at which the sharp peaks disappear and the system is fully localized, are extracted via a piecewise fit of the form $\text{max}\left(A\left(1-e^{(t-t_{coh})/\tau}\right),0\right)$ shown as lines in the insets. For increasing lattice depth, the time for which coherence persists drops, reaching only a few ms in the regime where the ground state would be localized across the entire cloud (red shading).\label{fig:Quench_phase_transition}}
\end{figure*}

\section{Long-term dynamics following slow quenches}
We finally investigate the dynamics following slower quenches from an initial superfluid state, focusing on how they can induce a loss of coherence in the cloud. 

To set the scene, we first measure the coherence properties of the system without any quench. We do this by loading into the quasicrystal in $45$ms followed by a $10$ms hold before imaging the atoms after the booster stage and a 9ms TOF. We extract the width of the central peak as in~\cite{BG2022} and plot the results in Fig.~\ref{fig:Quench_phase_transition}(a). For lattice depths below the critical depth for the non-interacting localization transition ($1.78E_r$) \cite{sbroscia2020observing}, the whole system remains delocalized, whilst above this lattice depth the system begins to localize. The weak but finite interactions at $10a_0$ have a delocalizing effect that increases with density. Accordingly, atoms at the edge of the cloud, where the density is lowest, will localize first, whilst the higher density centre of the cloud will remain coherent. Increasing the lattice depth further moves the boundary between localized and extended further inward towards the centre of the cloud. As a consequence, the sharp diffraction peaks observed in TOF reduce in amplitude until they disappear at approximately $3.2E_r$, when the centre of the cloud localizes, at which point the fitted width for the central peak jumps to that of the broad incoherent background cloud. As discussed in~\cite{BG2022}, the shrinking of the superfluid part of the cloud should in principle increase the widths of the diffraction peaks. This is however not observable in that paper or the present experiment, as the finite TOF means that in practice the widths of the superfluid peaks are dominated by the initial cloud size. As such we observe only a very sharp change in the measured cloud width at the point where coherence is fully lost, i.e., at the critical lattice depth for the density at the cloud centre.

To study the dynamics following slow quenches we first load to a lattice depth of $2E_r$, where only the outermost ring will be localized~\cite{BG2022}. This is followed by a slow, linear $290$\textmu s quench (ramp) to the final lattice depth and a variable hold time, before the booster stage and time-of-flight imaging. Fig.~\ref{fig:Quench_phase_transition}(b) shows the resulting momentum distributions for a final lattice depth of $2.6\,E_r$, where, in equilibrium, the centre of the cloud is superfluid. And indeed, the sharp superfluid peaks remain after the slow quench and show no dynamics on the sub-ms timescales observed after fast quenches, instead decaying gradually in amplitude over approximately $40\,$ms. We use a piecewise fit to the amplitude of the central diffraction peak to determine the critical hold time $t_{coh}$ after which the superfluid disappears and plot the results in Fig.~\ref{fig:Quench_phase_transition}(c) as a function of final lattice depth.

For quenches that end in the Bose glass regime ($V_{\text{latt}} \gtrsim  3.2 E_r$), the whole cloud localizes and coherence is mostly lost already during the ramp with any remaining coherence decaying rapidly on the timescale of a few ms. This is consistent with the result for fast quenches to the Bose glass regime (see Fig.\ref{fig:Quench_SF_BG}), where, for even deeper final lattices, the cloud decoheres within approximately $100$\textmu s. 

For quenches that end in the superfluid regime of $V_{\text{latt}} \lesssim  3.2\,E_r$, in contrast, the timescale of the coherence decay increases significantly for decreasing final lattice depths, suggesting a fundamentally different mechanism. We attribute this to the inhomogeneous nature of the system: During the quench, the denser central part of the cloud remains superfluid and the low-density outermost part remains localized. However, regions at intermediate densities will cross from the superfluid to the Bose glass. 

Atoms in the superfluid part of the cloud are able to respond quickly to changes in the system's Hamiltonian~\cite{BG2022} and as such will remain largely coherent, explaining the significant population of coherent peaks after the quench. However, the localized Bose glass is not able to respond adiabatically to changes in the Hamiltonian~\cite{BG2022}, resulting in significant entropy production in those regions that localize during the slow quench. Assuming that some of this entropy would then be transported from the inner edge of the localized region into the superfluid and thereby heat it up, this would naturally lead to the observed gradual decrease in the height of the diffraction peaks as more and more of the cloud decoheres.

Several conflicting factors come into play when estimating the time $t_{coh}$ for which superfluidity should persist. On the one hand, increasing the final lattice depth will shrink the region of the cloud that remains superfluid and hence decrease the distance over which entropy needs to be  transported to reach the trap centre. It will also reduce the amount of entropy required for the trap centre to decohere, as the superfluid approaches the critical lattice depth of the phase transition. Both of these effects will reduce $t_{coh}$. On the other hand, increasing the final lattice depth leads to slower transport, both due to localization effects and, more trivially, through a moderate reduction in all tunnelling amplitudes (though the average reduction in $\bar{J}_{\text{sum}}$ is only around a factor of 1.5 between $2.2\,E_r$ and $3\,E_r$). This slower transport should hence increase the time taken for entropy to reach the centre of the cloud and increase $t_{coh}$. Experimentally we find that overall the effects which reduce $t_{coh}$ win out, though the precise mechanisms at play behind our observations would be an interesting route for further study

\section{Conclusions}
We have experimentally studied the nonequilibrium dynamics of ultracold bosons following quantum quenches between the superfluid and Bose glass regimes in a 2D optical quasicrystal.  
For the quench to the BG regime, the dynamics are dominated by the spread in quasi-disordered on-site energies and can be described as a quantum walk in (4D) momentum space. By extracting the  trap-averaged real-space correlation function, we demonstrated that the quench to the superfluid regime in contrast is driven by tunneling and leads to an initially light-cone-like spreading of correlations, before the effects of the other terms in the hamiltonian lead to decoherence and, ultimately, thermalization. We could furthermore demonstrate that slow quenches within the partially superfluid regime give rise to a complex dynamics driven by entropy transport, opening the route to systematic studies of this central transport property, in particular close to localized phases and across interfaces between localized and delocalized regions. Our findings using 2D optical quasicrystals open many new avenues for experimentally exploring the non-equilibrium quantum dynamics of interacting disordered systems and many-body localization in two dimensions, in particular in regimes where all terms of the hamiltonian are relevant and classical simulations are unfeasible.

\begin{acknowledgments} This work was partly funded by the European Commission ERC Starting Grant \mbox{QUASICRYSTAL}, the EPSRC Grant EP/R044627/1 and Programme Grants \mbox{DesOEQ} (EP/P009565/1) and QQQS (EP/Y01510X/1) as well as the EPSRC QT Hub for Quantum Computing via Integrated and Interconnected Implementations (QCi3) (EP/Z53318X/1). B.S. acknowledges the National Natural Science Foundation of China (12374242) and the Beijing Natural Science Foundation (Z240007).\end{acknowledgments}

\appendix
\section{Further experimental details}\label{SuppMat:sequence}
The experimental sequence begins with preparing a Bose-Einstein condensate (BEC) of around $10^5$ $^{39}$K atoms in the $|F=1,m_F=1\rangle$ state in an optical dipole trap at $a_s\approx10\,a_0$. All lattices are formed using single-frequency laser beams with a 
wavelength $\lambda_{latt}\approx 725\,$nm. The lattice depth of the horizontal 2D quasicrystal (in-plane lattice)  and the vertical lattice are ramped up exponentially in $45\,$ms with a time constant of $10\,$ms, before the atoms are held in the initial lattice for an extra $10\,$ms. This initial part of the sequence is the same as that in~\cite{BG2022}.

For the quench measurements we always prepare the system at in-plane lattice depths of either $V_{\text{latt}}=6E_r$ (BG regime) or $V_{\text{latt}}=2E_r$ (SF regime), whereas for the static coherence measurement in Fig.~\ref{fig:Quench_phase_transition}(a) we load into a lattice of variable in-plane lattice depths up to $5E_r$. From here, we rapidly change the in-plane lattice depth in either $10\,$\textmu s or $290\,$\textmu s to induce a fast (as in Fig.~\ref{fig:Quench_SF_BG}) or a slow quench (as in Fig.~\ref{fig:Quench_phase_transition}(b)), respectively. The final state is then probed after a variable quench hold time at the final lattice depth. We omit these stages for the static coherence measurement.

\begin{figure}[h]
\centering
\includegraphics[width=0.48\textwidth]{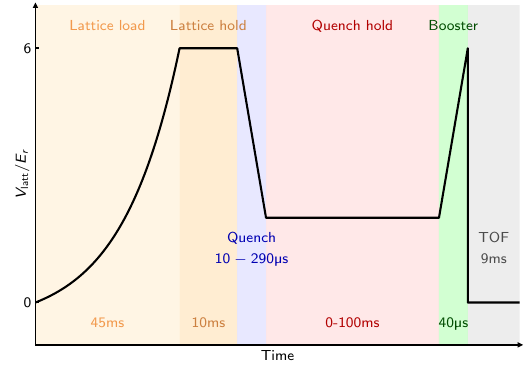}
\caption{\textbf{Experimental sequence for quantum quench measurements.} An example schematic showing the in-plane lattice depth throughout the different stages of a quench experiment - here from the BG regime to the SF regime. The quench itself is performed in either 10\textmu s or 290\textmu s for fast and slow quench experiments respectively, and this stage as well as the following hold time are entirely omitted for static coherence measurements. The booster stage is not included for fast quenches to the BG regime.}
\label{fig:FigS1}
\end{figure}

For the quench from the BG regime to the SF regime, the slow quench sequences, and the static coherence measurement, we implement a booster stage where we ramp the lattice depth to $6 E_r$ in $40\,$\textmu s just before the time-of-flight imaging. This redistributes the momentum population from the central peak to satellite peaks for easier extraction of the system's coherence. The ramp time is chosen to be fast enough compared to the tunneling dynamics so as not to significantly change the system's coherence. This method is inspired by ref.~\cite{stoferle2004transition} and was also used in \cite{BG2022}. We finally image using absorption imaging after a time-of-flight of $9\,$ms.

Note that due to limitations of our imaging setup, the OD we measure begins to saturate beyond a value of roughly 3. In this paper, this effect only affects images taken of the superfluid with no quench, or shortly after quenches from the superfluid regime, and is accordingly taken into account when extracting the relative momentum peak occupations after fast quenches for these times as discussed in appendix~\ref{SuppMat:high_order_peak_extraction}.

\begin{figure}[h]
    \centering
    \includegraphics[width=0.48\textwidth]{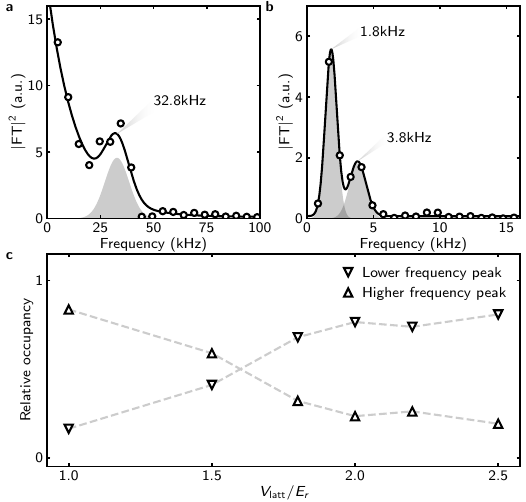}
    \caption{\textbf{Fourier analysis of quench dynamics.} Power spectra of $n(\mathbf{k}=\mathbf{0},t)$ for quenches from (a) $V_{\text{latt}}=2E_r$ to $V_{\text{latt}}=7E_r$ and (b) $V_{\text{latt}}=6E_r$ to $V_{\text{latt}}=1.8E_r$. Using fits (solid lines) to these data we extract the dominant frequencies and the relative power in the peaks using the area under each fitted peak (shaded grey). (c) The power in each peak normalized by the sum of the powers in both peaks plotted against the final lattice depth after the quench. \label{figS:FFT}}
\end{figure}

\section{Frequency extraction}\label{SuppMat:FreqExtraction}
To quantitatively analyze the nonequilibrium dynamics, we first bin the momentum distribution over momentum bins of width $0.17\hbar k_0$ before Fourier transforming the occupation of each bin in time using a fast Fourier transform (FFT). Since the FFT requires a constant time step, in cases where the time step between subsequent images varies we interpolate between images, using the smallest time step for that sequence of images. All relevant features in the time domain are well resolved already for the longest time step, and so are far from the Nyquist limit regardless of this interpolation.

Fig.~\ref{figS:FFT}(a) and (b) show two examples of the resulting power spectra ($|\text{FT}|^2$, where $\text{FT}$ is the Fourier transform) of the zero-momentum component $n(\mathbf{k}=\mathbf{0},t)$. For each quench, we extract the dominant frequencies by fitting to the power spectrum. For quenches from the SF to the BG regime, the spectra consist of a decaying function and a peak at a frequency $f_0$ and can be well fitted by:

\begin{equation}  \label{eq:SFtoBGQuenchFit}
    |\text{FT}(f)|^2 = Ae^{-f/L}+Be^{-(f-f_0)^2/2s^2}+C\,.
\end{equation}

For quenches from the BG to the SF regime, we fit with either a single Gaussian plus a constant, or, when more than one peak is visible, with a sum of two Gaussians plus a constant. In all cases we exclude the DC offset at $f=0$ in the fitting procedure.
\begin{figure}[h]
    \centering
    \includegraphics[width=0.48\textwidth]{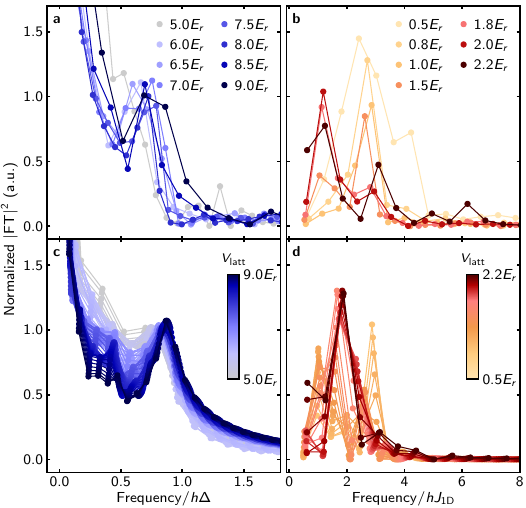}
    \caption{\textbf{Normalised quench dynamics.} The power spectrum at $\mathbf{k}=\mathbf{0}$ is plotted as a function of normalised frequencies. (a) and (c) show experimental and numerical power spectra (respectively) of dynamics after quenches from a SF ($V_{\text{latt}}=2E_r$ for experiment or $1.5E_r$ for theory) to the BG regime, where the frequency is normalised by the quasi-disorder strength $\Delta$. (b) and (d) show the equivalent for quenches from a BG ($V_{\text{latt}}=6E_r$) to different lattice depths in the SF regime. Here the frequency is normalised by the tunneling element $J_{1D}$ in a 1D periodic lattice of the same depth. In each case the spectra are normalized such that the sum of the heights of all fitted peaks is the same within the plot.\label{figS:Collapse}}
\end{figure}
Where two peaks are visible in the power spectrum, we extract their relative weights by using the area under the individual fitted peaks (shaded grey in Fig.~\ref{figS:FFT}) as a measure of the power in each peak. In Fig.~\ref{figS:FFT}(c) we plot these areas normalized by their sum and observe that while at lower lattice depths the higher frequency peak dominates, as the lattice gets deeper the lower frequency peak becomes more significant until it is eventually the dominant peak.

Fig.~\ref{figS:Collapse} shows the power spectra in dimensionless form where frequencies are renormalised using natural scales, namely the tunneling strength of a periodic 1D lattice $J_{1D}$ of the same depth $V_\text{latt}$ and the quasi-disorder strength $\Delta$ for the quenches to the SF and BG regimes respectively. In both cases, we find that the power spectra for each quench cover a similar range of normalized frequencies, suggesting that tunneling and quasi-disorder are indeed the relevant energy scales.

\begin{figure}[h]
    \centering
    \includegraphics[width=0.48\textwidth]{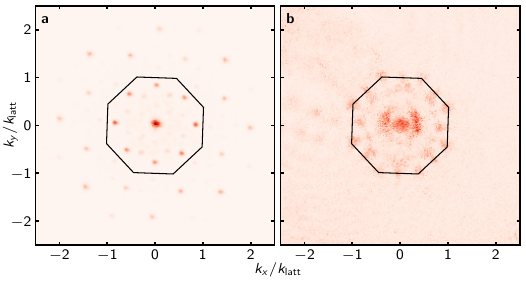}
    \caption{\textbf{Maximum square amplitude of the Fourier transform of the quench dynamics, with DC offset removed.}  (a) The quench from $V_{\text{latt}}=2E_r$ to $V_{\text{latt}}=7E_r$. The amplitude is strongly peaked at the low-order momentum peaks of the underlying quasicrystalline lattice. (b) The quench from $V_{\text{latt}}=6E_r$ to $V_{\text{latt}}=1E_r$. Notably, there are 16 peaks along the edge of the first pseudo Brillouin zone (PBZ, indicated by the black octagon in both images). These do not correspond to low-order momentum peaks. These images use the same dataset as Fig.~\ref{fig:Quench_SF_BG} in the main text, and are individually normalized. \label{figS:FFT_MaxInt}}
\end{figure}

Similar to the zero-momentum peak, we apply a time domain Fourier transform to the densities at all momenta $n(\mathbf{k},t)$ and plot the maximum square amplitude of the Fourier spectrum at each momentum in Fig.~\ref{figS:FFT_MaxInt}, excluding the DC offset. This indicates the parts of momentum space most involved in the quench dynamics, revealing another difference between quenches to different regimes. For the quench from the SF to the BG regime (Fig.~\ref{figS:FFT_MaxInt}(a)), the dynamics mainly involves momenta corresponding to low-order diffraction peaks of the underlying quasicrystalline lattice. For the quench from the BG to the SF regime (Fig.~\ref{figS:FFT_MaxInt}(b)), in contrast, different momenta such as the corners of the first pseudo Brillouin zone (PBZ) play an important part in the dynamics, consistent with Fig.~\ref{fig:Quench_SF_BG}.

\section{Simulations of quench dynamics} \label{SuppMat:theory}

\begin{figure*}
    \centering
    \includegraphics[width=\linewidth]{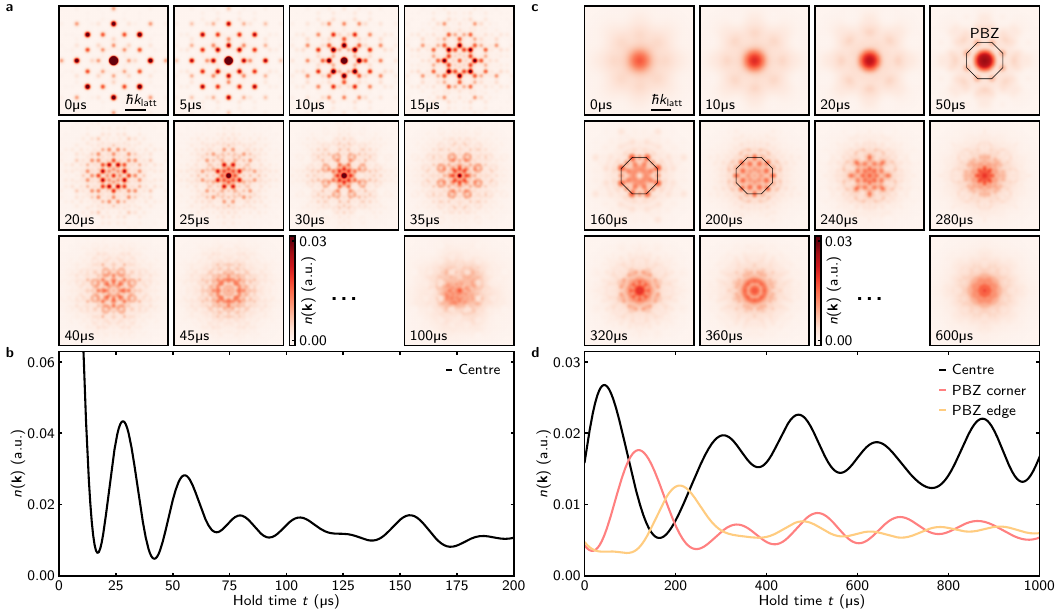}
    \caption{\textbf{Mean-field simulations of quench dynamics.} (a) and (c) show matter-wave dynamics after the quench from the SF ($V_{\text{latt}}=1.5E_r$) to the BG regime ($V_{\text{latt}}=7E_r$), and from the BG ($V_{\text{latt}}=6E_r$) to the SF regime ($V_{\text{latt}}=1E_r$) respectively. Images were convolved with a $0.1\hbar k_0$ wide Gaussian to account for experimental broadening (from a combination of imaging resolution, finite cloud size and finite TOF - see e.g.~\cite{gerbier_expansion_2008,doi:10.1073/pnas.1408861112}). (b) and (d) show populations of different momenta as a function of the hold time. Here the populations are normalised by the total number of atoms.\label{fig:Simulation}}
\end{figure*}

We simulate the dynamics of the quantum quenches using a discrete Hubbard model developed in~\cite{gottlobHubbardModelsQuasicrystalline2023}. It describes the lowest band of the quasicrystal, i.e., it contains one Wannier function $\ket{i}$ per lattice site $i$. Due to quasiperiodicity, all Hubbard parameters $J_{i,j}, \, U_i, \, \epsilon_i$ in Eq.~\ref{eq:Hubbard} are site-dependent and we calculate them numerically for each lattice depth $V_{\text{latt}}$ (the calculated parameters are accessible from~\cite{gottlob_schneider_2023}). We consider a circular patch of radius $12.5\,\lambda$ of the quasicrystal containing $n_s = 1622$ lattice sites with open boundary conditions.

For the quench from the superfluid to the Bose glass regime, we start from a Bose-condensed initial state where each atom is assumed to be in the (extended) non-interacting ground state for $V_i = 1.5E_r$, as obtained by exact diagonalisation:
\begin{equation}
    \ket{\Psi_{SF}(t=0)} = \sum_i c_i(t=0) \ket{i}\quad\text{with}\quad \sum_i |c_i|^2=1
\end{equation}

The dynamics is then simulated using a split-step mean-field approximation. For each time step $c_i(t)\rightarrow c_i(t+\delta t)$, we first time evolve with the non-interacting part of the Hamiltonian to get $\tilde{c}_i(t+\delta t)$, and then add the local mean-field phase shift due to on-site interactions (with $a_s = 10 a_0$):
\begin{equation}  \label{eq:MFinteraction}
    c_i(t+\delta t) =e^{-i \left(U_i n_{\text{atoms}}|c_i(t)|^2 \right)\delta t} \tilde{c}_i(t+\delta t)
\end{equation}
We assume on average unity filling such that the total atom number $n_{\text{atoms}}$ equals the number of lattice sites $n_s$ and use $\delta t = 0.625\,$\textmu s. The momentum space density is obtained using
\begin{align}
    n(\mathbf{k},t) &= \sum_{i,j}\bra{\Psi(t)}\hat{a}_i^{\dagger} \hat{a}_j\ket{\Psi(t)} \Tilde{w}_{i}^*(\mathbf{k})\Tilde{w}_{j}(\mathbf{k})\nonumber\\
    & = \sum_{i,j} c_i^{*}(t)c_j(t)\Tilde{w}_{i}^*(\mathbf{k})\Tilde{w}_{j}(\mathbf{k}) \nonumber \\
    & = \left|\sum_i c_i(t)  \Tilde{w}_{i}(\mathbf{k})\right|^2 \,,\label{eq:momentumdensity}
\end{align}
where $\Tilde{w}_{i}(\mathbf{k})$ is the Fourier transformed Wannier function at site $i$. Note that due to the quasi-periodic nature of the potential, the Wannier functions are distinct for each lattice site. In contrast to periodic lattices, where all Wannier functions are identical, it is convenient to explicitly work with the Wannier functions $w_i(\mathbf{r})$ instead of using a common shifted Wannier function $w(\mathbf{r}-\mathbf{r}_i)$. As a consequence, the often seen phase factors $\exp(i\mathbf{k}\cdot\mathbf{r}_i)$ are now included in $\Tilde{w}_{j}(\mathbf{k})$. The simulation includes the dynamics during the finite ramp time within which the simulated lattice depth is linearly raised from $V_i = 1.5E_r$ to $V_f$ in $10\,$\textmu s.

The initial lattice depth in the simulation differs from the one in the experiment and was chosen to provide a similar initial state, even though the simulation starts from a non-interacting ground state. While the single-particle ground state undergoes a localization transition at a lattice depth of $V_{loc} = 1.78E_r$~\cite{sbroscia2020observing}, the weakly-interacting initial state in the experiment is still extended, see Fig.~\ref{fig:Quench_phase_transition}. Appendix~\ref{SuppMat:InitialState} demonstrates that the dynamics depends only very weakly on the details of the initial state, justifying this approach (Fig.\ref{figS:InteractionAndInitial}).

The results of these simulations are shown in Fig.~\ref{fig:Simulation}(a,b) and show good qualitative agreement with the experiment. Differences in the initial state and the use of approximate mean-field dynamics preclude a direct side-by-side comparison. Nonetheless, the extracted timescales plotted in Figs.~\ref{fig:Quench_SF_BG}\&\ref{fig:Quench_SF_BG_freq} show good quantitative agreement, highlighting that these simulations capture the essential dynamics.

The quench from the Bose glass to the superfluid regime is modeled using the same split-step mean-field approach. We approximate the initial Hamiltonian at $V_i = 6E_r$ as being in the atomic limit and neglect tunneling. The simulation accordingly starts from a collection of fully-localized atoms. We consider a scattering length of $a = 10a_0$ and first calculate the initial on-site density profile $\{n_i\}$ of the ground state by tuning the chemical potential so that the average density across the lattice is unity. To avoid reflections of the particles on the open boundaries during the subsequent time evolution, we set the initial density to zero outside a circle with a radius of $6 \lambda$. 

Atoms originating from different lattice sites are initially in orthogonal states, and since the mean-field interaction considered during the simulation cannot entangle atoms, we simulate the wavefunctions originating from individual sites independently regarding the single-particle Hamiltonian ($c_i(t)\rightarrow c_i(t+\delta t)$). We then at each time step sum the densities from all individual wavefunctions to calculate the mean-field phase shifts according to Eq.~\ref{eq:MFinteraction} and apply them to each wavefunction. The final momentum density is then obtained by incoherently summing the momentum distributions (Eq.~\ref{eq:momentumdensity}) of particles originating on individual sites and weighting them by the initial densities $n_i$. Fig.~\ref{fig:Simulation}(c,d) shows the resulting dynamics and we again find good qualitative agreement with the experiment. In particular, the simulated dynamics also gives rise to transient momentum peaks at the corners and middle of the edges of the first pseudo-Brillouin zone (PBZ).

\section{Influence of the initial state and interactions on dynamics}\label{SuppMat:InitialState}

\begin{figure}[hbt]
    \centering
    \includegraphics[width=0.48\textwidth]{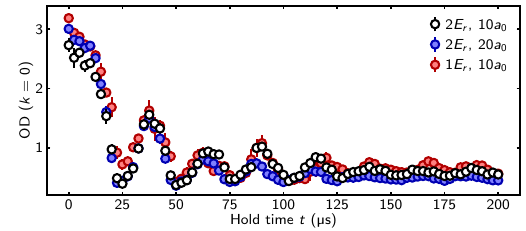}
    \caption{\textbf{Experimental quench dynamics for different initial lattice depths and interaction strengths.} For the quench from a delocalized SF to a deep lattice, both the precise initial state and the strength of the weak interaction play only a minor role as long as the initial state is delocalized. \label{figS:InteractionAndInitial}}
\end{figure}

In the weakly-interacting regime, we find that the dynamics after the quench from the SF to the BG regime depend only very weakly on the details of the initial state, as long as the system is quenched from an extended state. Fig.~\ref{figS:InteractionAndInitial} compares the dynamics between quenches from $V_{\text{latt}}=1E_r$ to $V_{\text{latt}}=7E_r$ (red circles) and from $V_{\text{latt}}=2E_r$ to $V_{\text{latt}}=7E_r$ (empty black circles, same data as Fig.~\ref{fig:Quench_SF_BG}(b)) with a scattering length of $a_s=10a_0$.

Although the initial system parameters are different, the dynamics are very similar, as are the extracted frequencies. We also explore the effect of interactions on the short-term dynamics by comparing to the dynamics from $V_{\text{latt}}=2E_r$ to $V_{\text{latt}}=7E_r$ at a scattering length of $a_s=20a_0$ (blue circles), demonstrating that for the weakly-interacting regime the short-term dynamics after the quench are indeed dominated by the final disorder strength.

\begin{figure}
    \centering
    \includegraphics[width=\linewidth]{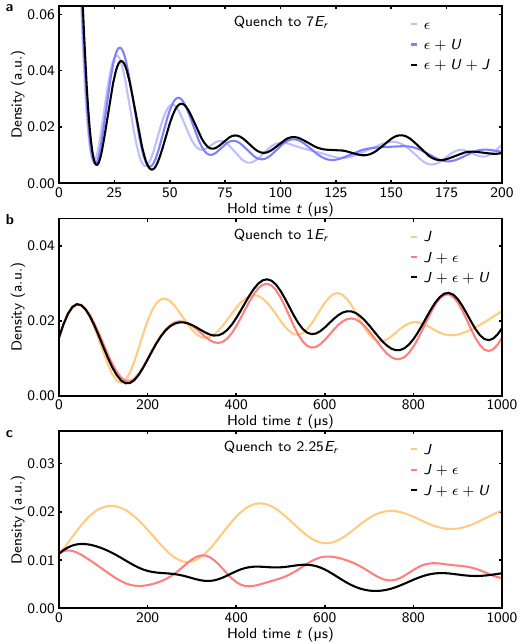}
    \caption{\textbf{Influence of the various Hubbard terms.} (a) The dynamics of the $\mathbf{k}=\mathbf{0}$ momentum peak for the quench from the superfluid ($V_{\text{latt}} = 1.0E_r$) to the Bose glass regime ($V_{\text{latt}} = 7E_r$). Here, on-site energies dominate while tunneling and mean-field interactions give rise to small corrections. (b) Quench from the Bose glass ($V_{\text{latt}} = 6E_r$) deep into the superfluid regime ($V_{\text{latt}} = 1E_r$). Here, tunneling dominates the dynamics. (c) When quenching to intermediate lattice depths ($V_{\text{latt}} = 6E_r$ to $V_{\text{latt}} = 2.25E_r$) close to the phase transition, all three Hubbard terms contribute to the dynamics.} \label{fig:comparison}
\end{figure}

Fig.~\ref{fig:comparison}(a) shows the simulated oscillation of $n(\mathbf{\mathbf{k} = \mathbf{0}},t)$ for a quench from $V_{\text{latt}} = 1.5E_r$ to $V_{\text{latt}} = 7E_r$. The discrete Hubbard model allows us to select which Hubbard parameters to include in the simulation, confirming that on-site energies dominate the quench dynamics from the superfluid to the Bose glass regime. Fig.~\ref{fig:comparison}(b) shows the simulated oscillation of the $\mathbf{k}=\mathbf{0}$ momentum peak for the quench from $V_{\text{latt}} = 6E_r$ to $V_{\text{latt}} = 1E_r$, with various subsets of the Hubbard parameters included, confirming that for shallow final lattice depths tunneling terms dominate the quench dynamics from the Bose glass to the superfluid regime. As shown in Fig.~\ref{fig:comparison} c, in contrast, the dynamics become more complex in quenches from the Bose glass to intermediate lattice depths, where the three Hubbard terms become comparable in size and hence all contribute significantly to the oscillation.

\begin{figure*}
    \centering
    \includegraphics{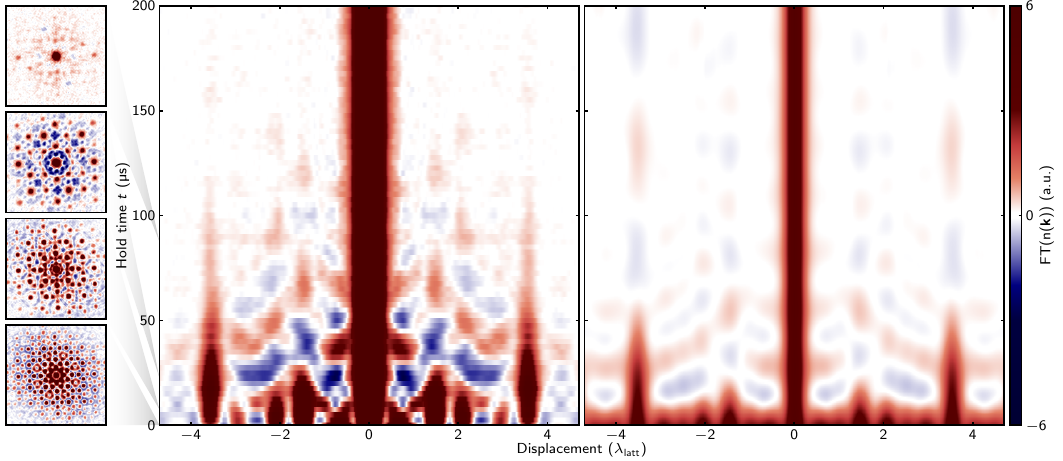}
    \caption{\textbf{Correlation decay in the quench from superfluid to Bose-glass regime.} Trap-averaged real-space correlations after a quench from $2E_r$ to $7E_r$. Small images on the left show measured correlations at selected times of 0\textmu s, 7.5\textmu s, 22.5\textmu s, and 87.5\textmu s.  Main panels show the average of cuts along the four lattice axes for both experiment, left, and mean-field simulation, right. Both clearly show the decay of correlations on all length scales caused by the quasiperiodic nature of the potential.}
    \label{fig:SpatialFFT_SFtoBG}
\end{figure*}

\section{Extracting trap-averaged real-space correlations}\label{SuppMat:real_space_autocorr}
A central observable in characterizing the coherence properties of a many-body bosonic quantum state $\ket{\Psi}$ is the first-order correlation function 
\[C_1(\mathbf{r}_1,\mathbf{r}_2)=\left\langle\hat{\psi}^\dag(\mathbf{r}_1)\hat{\psi}(\mathbf{r}_2)\right\rangle=\bra{\Psi}\hat{\psi}^\dag(\mathbf{r}_1)\hat{\psi}(\mathbf{r}_2)\ket{\Psi},\] 
where $\hat{\psi}(\mathbf{r})$ is the field annihilation operator at position $\mathbf{r}$.

The trap-averaged correlation function, 
\[g(\mathbf{r})=\int\left\langle\hat{\psi}^{\dag}(\mathbf{r}')\hat{\psi}(\mathbf{r}'+\mathbf{r})\right\rangle d\mathbf{r}',\]
is closely related to the momentum distribution and can hence be directly extracted from it without the need for a dedicated experimental apparatus such as a quantum gas microscope~\cite{murthy_observation_2015,boettcher_quasi-long-range_2016,murthy_quantum_2019,guo_observation_2024}.

Generally speaking, the autocorrelation $g(x)$ of a function f(x) is defined as
\[g(x)=\int f^*(x')f(x'+x)dx',\]
and, as stated by the Wiener-Khinchin theorem, it can be obtained from the function's power spectrum via an inverse Fourier transform (IFT): 
\[{g(x)=2\pi\,\text{IFT}[\left|\text{FT}[f(x)]\right|^2]}.\]
The prefactor $2\pi$ depends on the convention used to define the Fourier transform. Adapted to the case at hand, this becomes ${g(\mathbf{r})=2\pi\,\text{IFT}[n(\mathbf{k})]}$. 

More explicitly, for a quantum state $|\Psi\rangle$ expressed in the basis of Wannier functions $w_i$ using the usual on-site creation and annihilation operators $\hat{a}_i^{\dag}$ and $\hat{a}_i$, the momentum density is given by
\[n(\mathbf{k})= \sum_{i,j}\bra{\Psi} \hat{a}_i^{\dagger} \hat{a}_j\ket{\Psi} \Tilde{w}_{i}^*(\mathbf{k})\Tilde{w}_{j}(\mathbf{k}),\]
where $\Tilde{w}_{i}(\mathbf{k})=\text{FT}[{w}_{i}(\mathbf{r})]$. This leads to:
\begin{alignat}{3}
  \text{IFT}\left[ n(\mathbf{k}) \right]&=\sum_{i,j}\bra{\Psi}\hat{a}_i^{\dagger} \hat{a}_j\ket{\Psi} \text{IFT}\left[ \Tilde{w}_{i}^*(\mathbf{k})\Tilde{w}_{j}(\mathbf{k}) \right] \nonumber \\
   &=\sum_{i,j}\bra{\Psi}\hat{a}_i^{\dagger} \hat{a}_j\ket{\Psi}\times\nonumber\\
   &\quad\quad\text{IFT}\left[\int e^{i\mathbf{k}\cdot\mathbf{r'}}{w}_{i}^*(\mathbf{r}')d\mathbf{r}'\int e^{-i\mathbf{k}\cdot\mathbf{r}''}{w}_{j}(\mathbf{r}'')d\mathbf{r}''\right]\nonumber\\\
   &=\frac{1}{2\pi}\sum_{i,j}\bra{\Psi}\hat{a}_i^{\dagger} \hat{a}_j\ket{\Psi}\times\nonumber\\
   &\quad\quad\iiint e^{-i\mathbf{k}\cdot(\mathbf{r''}-\mathbf{r'}-\mathbf{r})}{w}_{i}^*(\mathbf{r}'){w}_{j}(\mathbf{r}'')d\mathbf{r}'d\mathbf{r}''d\mathbf{k}\nonumber\\
   &=\frac{1}{2\pi}\sum_{i,j}\bra{\Psi}\hat{a}_i^{\dagger} \hat{a}_j\ket{\Psi}\times\nonumber\\
   &\quad\quad\int{w}_{i}^*(\mathbf{r}'){w}_{j}(\mathbf{r}'+\mathbf{r})d\mathbf{r}'\nonumber\\
   &=\frac{1}{2\pi}g(\mathbf{r}).\label{eq:autocorrelation}
\end{alignat}

\begin{figure*}
    \centering
    \includegraphics{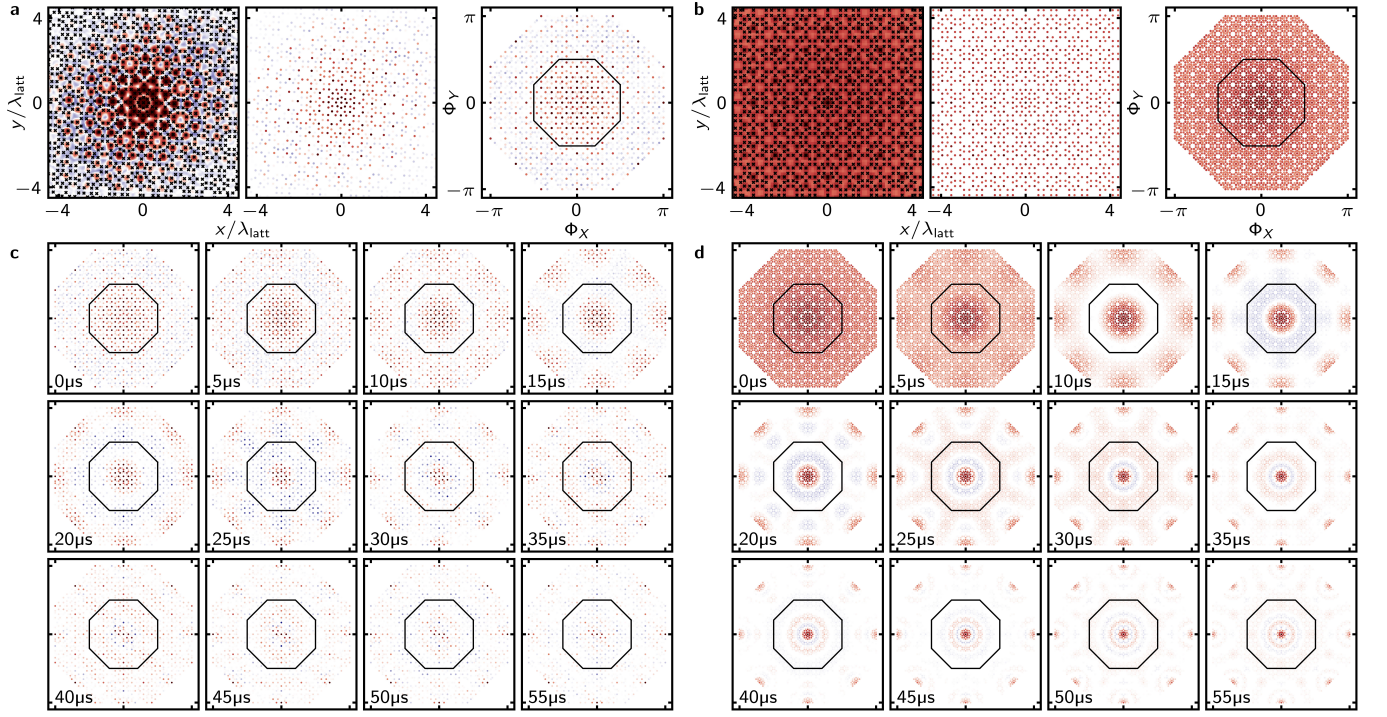}
    \caption{\textbf{Correlations in configuration space for SF to BG quench.} (a) Correlations for a quench from $2E_r$ to $7E_r$ for a hold duration of $0$\textmu s. The image on the left shows the measured trap-averaged correlations with small black crosses indicating the possible vectors between pairs of sites connected by up to 10 first-order hops along each lattice direction. The central image is a scatter plot of the correlation averaged within a radius of 0.1$\lambda_{\text{latt}}$ around each of the black crosses. The image on the right shows the same scatter plot in $O'$ with the black octagon indicating the original configuration space octagon $O$. (c) shows how the correlations in configuration space ($O'$) evolve with hold time after the quench (labels). (b) and (d) show the same for a numerical simulation of a quench from $1.5E_r$ to $7E_r$. We see qualitative agreement in the behaviours of experiment and theory (in spite of differences in the initial correlations), albeit with a time offset of order $5$\textmu s between them. The broad negative background visible at short times after the quench in the experimental trap-averaged correlations is caused by the saturation of the measured OD for the central peak $(\mathbf{k}=0)$ in momentum space. This saturation effectively acts as a negative peak centered at $\mathbf{k}=0$ in momentum space and hence becomes a broad negative background after Fourier transforming to real space. The color map is the same as that used in Fig.~\ref{fig:SpatialFFT_SFtoBG}.}
    \label{fig:AutocorrConfigurationSpace}
\end{figure*}
Note that in our case, where $n(\mathbf{k})$ is real and even such that $g(\mathbf{r})=g({-\mathbf{r}})$, taking the Fourier transform as opposed to the inverse Fourier transform gives the same result without the factor of $1/2\pi$, and this is what we use throughout this paper.

The integrals over products of Wannier functions ${{w}_{i}^*(\mathbf{r}'){w}_{j}(\mathbf{r}'+\mathbf{r})}$ imply that correlations are peaked at displacements $\mathbf{r}$ that correspond to the displacements between the maxima of two Wannier functions, i.e.\ the displacements between lattice sites, with the widths of the peaks determined by the widths of the Wannier functions. The coefficients $\bra{\Psi}\hat{a}_i^{\dagger} \hat{a}_j\ket{\Psi}$ in front of these integrals then weight each contribution by the correlation between these two sites in the state $\Psi$.

Fig.~\ref{fig:SpatialFFT_SFtoBG} shows the resulting trap-averaged correlations for a quench from the superfluid to the Bose-glass regime. The initial superfluid has significant long-range correlations, which gradually decrease with hold time due to the local dephasing caused by the quasi-disordered on-site energies. The finite experimental resolution in momentum space, caused by the combination of finite in-situ cloud size and finite time-of-flight~\cite{BG2022}, limits the observable correlations to around $r\lesssim6 \lambda$ even though the initial superfluid is expected to show (quasi)long-range coherence.

\section{Correlations in configuration space}

The structure in the decay pattern of the correlations becomes apparent by transforming it from real space to the configuration space described in \cite{gottlobHubbardModelsQuasicrystalline2023}. In configuration space sites are arranged within an octagon $O$ based on their shape, with the deepest sites at its centre, and the shallowest sites around its edges. We would hence expect that, given that the dynamics are governed largely by the on-site energies, this should show up as clear patterns in the correlation in configuration space. 
\begin{figure*}         
    \centering
    \includegraphics{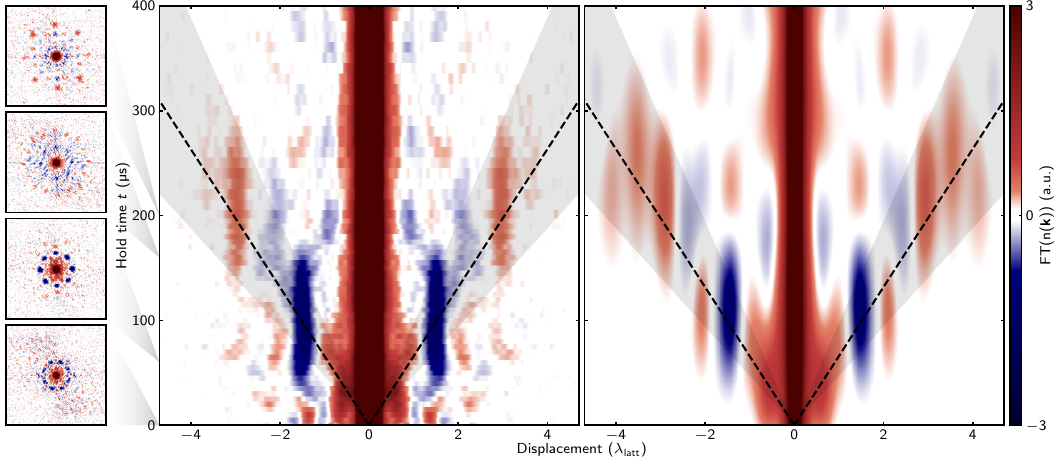}
    \caption{\textbf{Light-cone correlation spreading in the quench from Bose glass to superfluid regime.} Trap-averaged real-space correlations after a quench from $6E_r$ to $1E_r$. Small images on the left show measured correlations at selected times of 0\textmu s, 60\textmu s, 160\textmu s, and 200\textmu s.  Main panels show the average of cuts along the four lattice axes for both experiment, left, and mean-field simulation, right. Both exhibit a ``light cone" of correlations spreading over increasing length scales. This spreading of correlations is compatible with the ballistic expansion of initially localized atoms. Dashed lines follow $\text{displacement}=\pm2\bar{v}_{\text{max}}t$, and represent an estimate for the boundaries of the light-cone using the tight-binding model of the quasicrystal at a lattice depth of $V_{latt}=1E_r$. The shaded area corresponds to the range of velocities given by the distribution of ${v_{\text{max},\,i}=\text{max}\{|2J_{ij}d_{ij}/\hbar|:\ j=1,...,N_{\text{sites}}\}}$.}
    \label{fig:SpatialFFT_BGtoSF}
\end{figure*}

To transform the measured trap-averaged correlations to configuration space, we first find all possible vectors in configuration space connecting pairs of sites with up to 10 first-order `hops' along each lattice direction between them, where a hop is defined as connecting two first-order neighbours according to \cite{gottlobHubbardModelsQuasicrystalline2023}. These vectors represent the distances in configuration space between pairs of sites and hence fill an octagon $O'$ with twice the radius of the octagon $O$ containing the sites. For each such vector we then find the corresponding displacement in real space (see \cite{gottlobHubbardModelsQuasicrystalline2023} for details of the relationship between these spaces) by summing up the real-space displacements corresponding to each of the first-order hops that make up the vector in configuration space. Each of these first-order displacements has length ${(1+\sqrt{2})\lambda_{\text{latt}}/4}$, with those along X and Y having the same directions in real and configuration space, and those along T and D opposing directions in real and configuration space. We experimentally extract the trap-averaged correlation corresponding to each of these (higher-order) displacements by averaging the measured correlations within a radius of 0.1$\lambda_{\text{latt}}$ around it, see Fig~\ref{fig:AutocorrConfigurationSpace}(a). Overall, this process effectively maps the correlation in real space to configuration space (from parallel space to perpendicular space in the cut and project scheme~\cite{walter_crystallography_2009}).

The resulting correlations in configuration space are plotted in Fig~\ref{fig:AutocorrConfigurationSpace}(c) and can explained as follows. The centre of $O'$ corresponds to sites that are very close in configuration space and hence have very similar on-site energies and therefore very similar phase evolutions. Moving outwards from the centre of $O'$, the distance between sites in configuration space initially gets larger. Hence they have on average a larger difference in their on-site energies, and as such the phase difference between the sites will increase faster. This leads to a series of concentric rings of positive and negative correlation around the origin of $O'$, whose radii become smaller with time.

For very large distances in configuration space, however, the behavior changes as, for the largest distances in $O'$  the trap-averaged correlation function only has contributions from sites close to opposite edges of the octagon $O$. Due to the symmetry of the octagon, these again have rather similar energies so their differential phase evolution is slower.
Hence, looking in configuration space, we also see a series of concentric rings appearing around the middles of the edges of $O'$ as time goes on, similar to those around the origin.

\section{Correlations in BG to SF quench}

In the case of the quench from the Bose glass to the superfluid regime (Fig~\ref{fig:SpatialFFT_BGtoSF}), the trap-averaged correlation function displays a radically different behaviour. Here, the strongly localized initial state possesses only very short-range correlations, but the dynamics gives rise to a light-cone-like structure caused by the spreading of correlation over increasingly large length scales, compatible with a ballistic spreading of the initially localized particles~\cite{Cheneau2012}. The locations of the emerging correlation peaks agree well between experiment and simulation and again directly correspond to the possible displacements between sites in the quasicrystal.

We can estimate the gradient of the cone in analogy to periodic lattices: in the periodic case the maximum group velocity in the lowest band is given by twice the tunneling rate $J/\hbar$ multiplied by the distance between sites $d$. Considering a particle initially localized on a single site, correlated wavepackets will expand in each direction at maximum velocity $2Jd/\hbar$ \cite{schneider_fermionic_2012}, and hence correlations will appear at maximum displacements $\pm 4Jdt/\hbar$ after time $t$. Extending this reasoning to the quasicrystal, we calculate the quantity

\begin{equation}
    \bar{v}_{\text{max}}=\frac{1}{N_{\text{sites}}}\sum_i \text{max}\{|2J_{ij}d_{ij}/\hbar|:\ j=1,...,N_{\text{sites}}\}
\end{equation}
where $d_{ij}$ is the real-space distance between sites $i$ and $j$, and $N_{\text{sites}}$ is the number of sites in the system. This can be seen as the equivalent of the maximum group velocity based on the tunneling elements and distances between individual sites, averaged across the whole lattice. The light-cone predicted by this velocity ($\text{displacement}=\pm2\bar{v}_{\text{max}}t$) is plotted as dashed lines in Figs.~\ref{fig:Quench_SF_BG} and~\ref{fig:SpatialFFT_BGtoSF}, and matches the light cone seen in the experimental data well.

For late times, these transient correlations decay again as the other terms in the hamiltonian decohere the correlations and ultimately lead to thermalization into a hot thermal state.

\section{Extracting peak populations up to high orders}\label{SuppMat:high_order_peak_extraction}
In order to extract the population of each momentum peak in the fast SF to BG quenches, we employ the following fitting procedure. We start by extracting the eight fundamental momentum space vectors along the eight lattice directions and the location of the $\mathbf{k}=\mathbf{0}$ point using Gaussian fits to the eight first-order peaks. Here we average our result over images with $t\leq 10$\textmu s, where these peaks are clear and easy to separate from any higher order peaks, and also average the fitted standard deviations along the $X$ and $Y$ directions for later use - we will refer to these averages as $\sigma_{X,Y}$ respectively.
\begin{figure}[h!]
    \centering
    \includegraphics{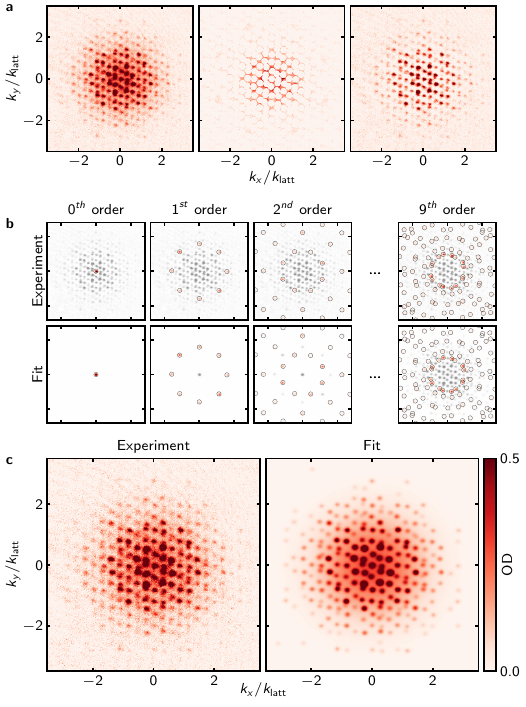}
    \caption{\textbf{Extracting momentum peak populations.} (a) Starting from the raw image (left) we first create a masked version (middle), where all peaks are masked out. We then fit the broad background and subtract it (right). (b) Peak fitting procedure: starting from 0th order (left), we sequentially go through each momentum order up to 9 (from left to right). Each peak is fitted individually by a Gaussian within a circular region of interest (black circles, radius $0.16\hbar k_0$). The fitted Gaussian is then subtracted from the image before moving on to the next peak. Red indicates the currently fitted momentum order (top row) and the fit result (bottom). In the bottom row, grey indicates the already fitted peaks - by including higher and higher orders, we build up a better fit of the overall image. (c) Raw image (left), and final fitted version including the fitted background (right).}
    \label{fig:PeakExtractionSchematic}
\end{figure}

Then, as shown in Fig.\ref{fig:PeakExtractionSchematic}(a), for each individual image we initially mask out circles around the expected peak locations and fit the remainder of the image with a broad Gaussian to extract the density distribution of atoms in the background cloud, which will consist of both incoherent atoms and atoms in peaks which are of orders higher than those we consider. After subtracting this background, we fit the peaks one by one, after each step subtracting the fitted Gaussian from the image before proceeding to the next fit. This ensures that for closely spaced peaks, the tails of one peak do not show up as a false signal at the neighboring fit. The only free fit parameters are the amplitude of the peak and a constant background value (both of which are fitted separately for each peak), the peak positions are calculated from the fundamental vectors extracted above and the widths are fixed to the $\sigma_{X,Y}$ of the initial fits to the $t\leq 10$\textmu s images. Fig.\ref{fig:PeakExtractionSchematic}(b) illustrates this fitting procedure from which we extract the population of each momentum peak. We set all populations where the peak height is less than OD 0.05 to zero, to prevent erroneous measured populations (for instance from fits capturing background noise or fringes induced by the imaging system, instead of atom density) from affecting our results significantly. From these peak populations we can calculate the RMS, shown in Fig.~\ref{fig:Quench_SF_BG} (d), in the 4D hypercubic lattice from which the quasicrystal's momentum space is projected from, see~\cite{viebahn2019matter} for details.

For short times after the quench, the population of the $\mathbf{k}=\mathbf{0}$ peak is very large, which poses an issue as our imaging setup cannot reliably measure OD values above approximately 3. Because we cannot then fit this peak reliably, its tails would not get properly removed when we subtract the fitted OD, which can lead to artificially higher fitted populations for high order peaks close to $\mathbf{k}=\mathbf{0}$. To get around this issue, for images where more than a quarter of the pixels in the region that would be fitted to for the central peak have an OD above 2.5, we first mask out a larger region for the central peak when fitting the broad background. When then fitting the central peak, we only fit to its tails, and then set in the image all pixels within the region we initially masked out for this peak to zero before moving on to any other fits (this will also set the populations of high order peaks close to the centre to zero, but these are regardless not significantly populated at the very short times where this issue is relevant). This procedure effectively eliminates the effect of this large central peak from the fits to the other peaks, but does not guarantee that the fit to its amplitude is accurate: as such for times where the fitted $\mathbf{k}=\mathbf{0}$ peak has a height greater than OD 3 (corresponding to $t\leq 15$\textmu s), we set the $\mathbf{k}=\mathbf{0}$ population such that the total population matches the mean total population of all other times.

\bibliography{BoseGlassQuench.bib}

@article{Cheneau2012,
  title={Light-cone-like spreading of correlations in a quantum many-body system},
  author={Cheneau, Marc and Barmettler, Peter and Poletti, Dario and Endres, Manuel and Schau{\ss}, Peter and Fukuhara, Takeshi and Gross, Christian and Bloch, Immanuel and Kollath, Corinna and Kuhr, Stefan},
  journal={Nature},
  volume={481},
  number={7382},
  pages={484--487},
  year={2012},
  publisher={Nature Publishing Group UK London},
  doi={10.1038/nature10748}
}

@article{koehn2026,
  title = {Quantum-gas microscopy and Talbot interferometry of the Bose-glass phase},
  author = {Koehn, Lennart and Parsonage, Christopher and Duncan, Callum W. and Kirton, Peter and Daley, Andrew J. and Hilker, Timon and Haller, Elmar and La Rooij, Arthur and Kuhr, Stefan},
  journal = {Phys. Rev. A},
  volume = {113},
  issue = {4},
  pages = {043303},
  numpages = {9},
  year = {2026},
  month = {Apr},
  publisher = {American Physical Society},
  doi = {10.1103/xb42-j6px},
  url = {https://link.aps.org/doi/10.1103/xb42-j6px}
}

@article{schreiber_observation_2015,
	title = {Observation of many-body localization of interacting fermions in a quasirandom optical lattice},
	volume = {349},
	url = {https://www.science.org/doi/full/10.1126/science.aaa7432},
	doi = {10.1126/science.aaa7432},
	pages = {842--845},
	number = {6250},
	journal = {Science},
	author = {Schreiber, Michael and Hodgman, Sean S. and Bordia, Pranjal and Lüschen, Henrik P. and Fischer, Mark H. and Vosk, Ronen and Altman, Ehud and Schneider, Ulrich and Bloch, Immanuel},
	urldate = {2024-02-21},
	year = {2015},
    month={08},
    day={21},
}

@article{lieb_finite_1972,
	title = {The finite group velocity of quantum spin systems},
	volume = {28},
	issn = {1432-0916},
	url = {https://doi.org/10.1007/BF01645779},
	doi = {10.1007/BF01645779},
	pages = {251--257},
	number = {3},
	journal = {Communications in Mathematical Physics},
	shortjournal = {Commun.Math. Phys.},
	author = {Lieb, Elliott H. and Robinson, Derek W.},
	urldate = {2026-09-04},
	year = {1972},
    month={09},
    day={01},
	langid = {english},
}

@article{schneider_fermionic_2012,
	title = {Fermionic transport and out-of-equilibrium dynamics in a homogeneous Hubbard model with ultracold atoms},
	volume = {8},
	rights = {2012 Springer Nature Limited},
	issn = {1745-2481},
	url = {https://www.nature.com/articles/nphys2205},
	doi = {10.1038/nphys2205},
	pages = {213--218},
	number = {3},
	journal = {Nature Physics},
	shortjournal = {Nature Phys},
	author = {Schneider, Ulrich and Hackermüller, Lucia and Ronzheimer, Jens Philipp and Will, Sebastian and Braun, Simon and Best, Thorsten and Bloch, Immanuel and Demler, Eugene and Mandt, Stephan and Rasch, David and Rosch, Achim},
	urldate = {2024-11-01},
	year = {2012},
    month={03},
	langid = {english},
}

@article{rigol2008thermalization,
  title={Thermalization and its mechanism for generic isolated quantum systems},
  author={Rigol, Marcos and Dunjko, Vanja and Olshanii, Maxim},
  journal={Nature},
  volume={452},
  number={7189},
  pages={854--858},
  year={2008},
  publisher={Nature Publishing Group},
  doi = {10.1038/nature06838}
}

@article{gottlobHubbardModelsQuasicrystalline2023,
  title = {Hubbard Models for Quasicrystalline Potentials},
  author = {Gottlob, E. and Schneider, U.},
  year = {2023},
  month = apr,
  journal = {Physical Review B},
  volume = {107},
  number = {14},
  pages = {144202},
  doi = {10.1103/PhysRevB.107.144202},
  urldate = {2023-11-26},
}

@article{eisert2015quantum,
  title={Quantum many-body systems out of equilibrium},
  author={Eisert, Jens and Friesdorf, Mathis and Gogolin, Christian},
  journal={Nature Physics},
  volume={11},
  number={2},
  pages={124--130},
  year={2015},
  publisher={Nature Publishing Group},
  doi = {10.1038/nphys3215}
}

@article{bernien2017probing,
  title={Probing many-body dynamics on a 51-atom quantum simulator},
  author={Bernien, Hannes and Schwartz, Sylvain and Keesling, Alexander and Levine, Harry and Omran, Ahmed and Pichler, Hannes and Choi, Soonwon and Zibrov, Alexander S and Endres, Manuel and Greiner, Markus and others},
  journal={Nature},
  volume={551},
  number={7682},
  pages={579--584},
  year={2017},
  publisher={Nature Publishing Group},
  doi ={10.1038/nature24622}
}

@article{zhang2017observation,
  title={Observation of a many-body dynamical phase transition with a 53-qubit quantum simulator},
  author={Zhang, Jiehang and Pagano, Guido and Hess, Paul W and Kyprianidis, Antonis and Becker, Patrick and Kaplan, Harvey and Gorshkov, Alexey V and Gong, Z-X and Monroe, Christopher},
  journal={Nature},
  volume={551},
  number={7682},
  pages={601--604},
  year={2017},
  publisher={Nature Publishing Group},
  doi = {10.1038/nature24654}
}

@article{choi2016exploring,
  title={Exploring the many-body localization transition in two dimensions},
  author={Choi, Jae-yoon and Hild, Sebastian and Zeiher, Johannes and Schau{\ss}, Peter and Rubio-Abadal, Antonio and Yefsah, Tarik and Khemani, Vedika and Huse, David A and Bloch, Immanuel and Gross, Christian},
  journal={Science},
  volume={352},
  number={6293},
  pages={1547--1552},
  year={2016},
  publisher={American Association for the Advancement of Science},
  doi = {10.1126/science.aaf8834}
}

@article{zhang2017observationTimeCrystal,
  title={Observation of a discrete time crystal},
  author={Zhang, Jiehang and Hess, Paul W and Kyprianidis, A and Becker, Petra and Lee, A and Smith, J and Pagano, Gaetano and Potirniche, I-D and Potter, Andrew C and Vishwanath, Ashvin and others},
  journal={Nature},
  volume={543},
  number={7644},
  pages={217--220},
  year={2017},
  publisher={Nature Publishing Group},
  doi = {10.1038/nature21413}
}

@article{choi2017observation,
  title={Observation of discrete time-crystalline order in a disordered dipolar many-body system},
  author={Choi, Soonwon and Choi, Joonhee and Landig, Renate and Kucsko, Georg and Zhou, Hengyun and Isoya, Junichi and Jelezko, Fedor and Onoda, Shinobu and Sumiya, Hitoshi and Khemani, Vedika and others},
  journal={Nature},
  volume={543},
  number={7644},
  pages={221--225},
  year={2017},
  publisher={Nature Publishing Group},
  doi = {10.1038/nature21426}
}

@article{greiner2002collapse,
  title={Collapse and revival of the matter wave field of a Bose--Einstein condensate},
  author={Greiner, Markus and Mandel, Olaf and H{\"a}nsch, Theodor W and Bloch, Immanuel},
  journal={Nature},
  volume={419},
  number={6902},
  pages={51--54},
  year={2002},
  publisher={Nature Publishing Group},
  doi = {10.1038/nature00968}
}

@article{will2010time,
  title={Time-resolved observation of coherent multi-body interactions in quantum phase revivals},
  author={Will, Sebastian and Best, Thorsten and Schneider, Ulrich and Hackerm{\"u}ller, Lucia and L{\"u}hmann, Dirk-S{\"o}ren and Bloch, Immanuel},
  journal={Nature},
  volume={465},
  number={7295},
  pages={197--201},
  year={2010},
  publisher={Nature Publishing Group},
  doi = {10.1038/nature09036}
}

@book{abrahams201050,
  title={50 years of Anderson Localization},
  author={Abrahams, Elihu},
  volume={24},
  year={2010},
  publisher={World Scientific},
  doi = {10.1142/7663}
}

@article{anderson1958absence,
  title={Absence of diffusion in certain random lattices},
  author={Anderson, Philip W},
  journal={Physical Review},
  volume={109},
  number={5},
  pages={1492},
  year={1958},
  publisher={APS},
  doi = {10.1103/PhysRev.109.1492}
}

@article{giamarchi1988anderson,
  title={Anderson localization and interactions in one-dimensional metals},
  author={Giamarchi, Thierry and Schulz, HJ},
  journal={Physical Review B},
  volume={37},
  number={1},
  pages={325},
  year={1988},
  publisher={APS},
  doi={10.1103/PhysRevB.37.325}
}

@article{fisher1989boson,
  title={Boson localization and the superfluid-insulator transition},
  author={Fisher, Matthew PA and Weichman, Peter B and Grinstein, Geoffrey and Fisher, Daniel S},
  journal={Physical Review B},
  volume={40},
  number={1},
  pages={546},
  year={1989},
  publisher={APS},
  doi={10.1103/PhysRevB.40.546}
}

@article{fallani2007ultracold,
  title={Ultracold atoms in a disordered crystal of light: Towards a Bose glass},
  author={Fallani, L and Lye, JE and Guarrera, V and Fort, C and Inguscio, M},
  journal={Physical Review Letters},
  volume={98},
  number={13},
  pages={130404},
  year={2007},
  publisher={APS},
  doi={10.1103/PhysRevLett.98.130404}
}

@article{d2014observation,
  title = {Observation of a Disordered Bosonic Insulator from Weak to Strong Interactions},
  author = {D'Errico, Chiara and Lucioni, Eleonora and Tanzi, Luca and Gori, Lorenzo and Roux, Guillaume and McCulloch, Ian P. and Giamarchi, Thierry and Inguscio, Massimo and Modugno, Giovanni},
  journal = {Physical Review Letters},
  volume = {113},
  issue = {9},
  pages = {095301},
  numpages = {5},
  year = {2014},
  month = {Aug},
  publisher = {American Physical Society},
  doi = {10.1103/PhysRevLett.113.095301},
  url = {https://link.aps.org/doi/10.1103/PhysRevLett.113.095301}
}

@article{gadway2011glassy,
  title={Glassy behavior in a binary atomic mixture},
  author={Gadway, Bryce and Pertot, Daniel and Reeves, Jeremy and Vogt, Matthias and Schneble, Dominik},
  journal={Physical Review Letters},
  volume={107},
  number={14},
  pages={145306},
  year={2011},
  publisher={APS},
  doi={10.1103/PhysRevLett.107.145306}
}

@article{pasienski2010disordered,
  title={A disordered insulator in an optical lattice},
  author={Pasienski, Matt and McKay, David and White, Matt and DeMarco, Brian},
  journal={Nature Physics},
  volume={6},
  number={9},
  pages={677--680},
  year={2010},
  publisher={Nature Publishing Group},
  doi={10.1038/nphys1726}
}

@article{meldgin2016probing,
  title={Probing the Bose glass--superfluid transition using quantum quenches of disorder},
  author={Meldgin, Carolyn and Ray, Ushnish and Russ, Philip and Chen, David and Ceperley, David M and DeMarco, Brian},
  journal={Nature Physics},
  volume={12},
  number={7},
  pages={646--649},
  year={2016},
  publisher={Nature Publishing Group},
  doi={10.1038/nphys3695}
}

@article{yang2017dynamical,
  title={Dynamical signature of localization-delocalization transition in a one-dimensional incommensurate lattice},
  author={Yang, Chao and Wang, Yucheng and Wang, Pei and Gao, Xianlong and Chen, Shu},
  journal={Physical Review B},
  volume={95},
  number={18},
  pages={184201},
  year={2017},
  publisher={APS},
  doi={10.1103/PhysRevB.95.184201}
}

@article{viebahn2019matter,
  title={Matter-wave diffraction from a quasicrystalline optical lattice},
  author={Viebahn, Konrad and Sbroscia, Matteo and Carter, Edward and Yu, Jr-Chiun and Schneider, Ulrich},
  journal={Physical Review Letters},
  volume={122},
  number={11},
  pages={110404},
  year={2019},
  publisher={APS},
  doi={10.1103/PhysRevLett.122.110404}
}

@article{sbroscia2020observing,
  title={Observing Localization in a 2D Quasicrystalline Optical Lattice},
  author={Sbroscia, Matteo and Viebahn, Konrad and Carter, Edward and Yu, Jr-Chiun and Gaunt, Alexander and Schneider, Ulrich},
  journal={Physical Review Letters},
  volume={125},
  number={20},
  pages={200604},
  year={2020},
  publisher={APS},
  doi={10.1103/PhysRevLett.125.200604}
}

@article{billy2008direct,
  title={Direct observation of Anderson localization of matter waves in a controlled disorder},
  author={Billy, Juliette and Josse, Vincent and Zuo, Zhanchun and Bernard, Alain and Hambrecht, Ben and Lugan, Pierre and Cl{\'e}ment, David and Sanchez-Palencia, Laurent and Bouyer, Philippe and Aspect, Alain},
  journal={Nature},
  volume={453},
  number={7197},
  pages={891--894},
  year={2008},
  publisher={Nature Publishing Group},
  doi={10.1038/nature07000}
}

@article{roati2008anderson,
  title={Anderson localization of a non-interacting Bose--Einstein condensate},
  author={Roati, Giacomo and D'Errico, Chiara and Fallani, Leonardo and Fattori, Marco and Fort, Chiara and Zaccanti, Matteo and Modugno, Giovanni and Modugno, Michele and Inguscio, Massimo},
  journal={Nature},
  volume={453},
  number={7197},
  pages={895--898},
  year={2008},
  publisher={Nature Publishing Group},
  doi={10.1038/nature07071}
}

@article{kondov2011three,
  title={Three-dimensional Anderson localization of ultracold matter},
  author={Kondov, SS and McGehee, WR and Zirbel, JJ and DeMarco, B},
  journal={Science},
  volume={334},
  number={6052},
  pages={66--68},
  year={2011},
  publisher={American Association for the Advancement of Science},
  doi={10.1126/science.1209019}
}

@article{jendrzejewski2012three,
  title={Three-dimensional localization of ultracold atoms in an optical disordered potential},
  author={Jendrzejewski, Fred and Bernard, Alain and Mueller, Killian and Cheinet, Patrick and Josse, Vincent and Piraud, Marie and Pezz{\'e}, Luca and Sanchez-Palencia, Laurent and Aspect, Alain and Bouyer, Philippe},
  journal={Nature Physics},
  volume={8},
  number={5},
  pages={398--403},
  year={2012},
  publisher={Nature Publishing Group},
  doi={10.1038/nphys2256}
}

@article{BG2022,
	title = {Observing the two-dimensional {Bose} glass in an optical quasicrystal},
	volume = {633},
	copyright = {2024 The Author(s)},
	issn = {1476-4687},
	url = {https://www.nature.com/articles/s41586-024-07875-2},
	doi = {10.1038/s41586-024-07875-2},
	number = {8029},
	urldate = {2024-09-25},
	journal = {Nature},
	author = {Yu, Jr-Chiun and Bhave, Shaurya and Reeve, Lee and Song, Bo and Schneider, Ulrich},
	month = sep,
	year = {2024},
	pages = {338--343},
}

@article{abanin2019colloquium,
  title={Colloquium: Many-body localization, thermalization, and entanglement},
  author={Abanin, Dmitry A and Altman, Ehud and Bloch, Immanuel and Serbyn, Maksym},
  journal={Reviews of Modern Physics},
  volume={91},
  number={2},
  pages={021001},
  year={2019},
  publisher={APS},
  doi = {10.1103/RevModPhys.91.021001}
}

@article{levi2012hypertransport,
  title={Hyper-transport of light and stochastic acceleration by evolving disorder},
  author={Levi, Liad and Krivolapov, Yevgeny and Fishman, Shmuel and Segev, Mordechai  },
  journal={Nature Physics},
  volume={8},
  pages={912--917},
  year={2012},
  publisher={Nature Publishing Group},
  doi = {10.1038/nphys2463}
}

@article{stoferle2004transition,
  title={Transition from a strongly interacting 1D superfluid to a Mott insulator},
  author={St{\"o}ferle, Thilo and Moritz, Henning and Schori, Christian and K{\"o}hl, Michael and Esslinger, Tilman},
  journal={Physical Review Letters},
  volume={92},
  number={13},
  pages={130403},
  year={2004},
  publisher={APS},
  doi = {10.1103/PhysRevLett.92.130403}
}

@article{banerjee_atomic_2012,
	title = {Atomic {Quantum} {Simulation} of {Dynamical} {Gauge} {Fields} {Coupled} to {Fermionic} {Matter}: {From} {String} {Breaking} to {Evolution} after a {Quench}},
	volume = {109},
	shorttitle = {Atomic {Quantum} {Simulation} of {Dynamical} {Gauge} {Fields} {Coupled} to {Fermionic} {Matter}},
	url = {https://link.aps.org/doi/10.1103/PhysRevLett.109.175302},
	doi = {10.1103/PhysRevLett.109.175302},
	number = {17},
	urldate = {2024-07-16},
	journal = {Physical Review Letters},
	author = {Banerjee, D. and Dalmonte, M. and M{\"u}ller, M. and Rico, E. and Stebler, P. and Wiese, U.-J. and Zoller, P.},
	month = oct,
	year = {2012},
	pages = {175302},
}

@article{marcos_two-dimensional_2014,
	title = {Two-dimensional lattice gauge theories with superconducting quantum circuits},
	volume = {351},
	issn = {0003-4916},
	url = {https://www.sciencedirect.com/science/article/pii/S0003491614002711},
	doi = {10.1016/j.aop.2014.09.011},
	urldate = {2024-07-16},
	journal = {Annals of Physics},
	author = {Marcos, D. and Widmer, P. and Rico, E. and Hafezi, M. and Rabl, P. and Wiese, U. -J. and Zoller, P.},
	month = dec,
	year = {2014},
	pages = {634--654},
}

@article{barbiero_coupling_2019,
	title = {Coupling ultracold matter to dynamical gauge fields in optical lattices: {From} flux attachment to $\mathbb{Z}_2$ lattice gauge theories},
	volume = {5},
	shorttitle = {Coupling ultracold matter to dynamical gauge fields in optical lattices},
	url = {https://www.science.org/doi/10.1126/sciadv.aav7444},
	doi = {10.1126/sciadv.aav7444},
	number = {10},
	urldate = {2024-07-16},
	journal = {Science Advances},
	author = {Barbiero, Luca and Schweizer, Christian and Aidelsburger, Monika and Demler, Eugene and Goldman, Nathan and Grusdt, Fabian},
	month = oct,
	year = {2019},
	pages = {eaav7444},
}

@article{arguello-luengo_engineering_2021,
	title = {Engineering analog quantum chemistry {Hamiltonians} using cold atoms in optical lattices},
	volume = {103},
	url = {https://link.aps.org/doi/10.1103/PhysRevA.103.043318},
	doi = {10.1103/PhysRevA.103.043318},
	number = {4},
	urldate = {2024-07-16},
	journal = {Physical Review A},
	author = {Arg{\"u}ello-Luengo, Javier and Shi, Tao and Gonz{\'a}lez-Tudela, Alejandro},
	month = apr,
	year = {2021},
	pages = {043318},
}

@article{lu_simulation_2011,
	title = {Simulation of {Chemical} {Isomerization} {Reaction} {Dynamics} on a {NMR} {Quantum} {Simulator}},
	volume = {107},
	url = {https://link.aps.org/doi/10.1103/PhysRevLett.107.020501},
	doi = {10.1103/PhysRevLett.107.020501},
	number = {2},
	urldate = {2024-07-16},
	journal = {Physical Review Letters},
	author = {Lu, Dawei and Xu, Nanyang and Xu, Ruixue and Chen, Hongwei and Gong, Jiangbin and Peng, Xinhua and Du, Jiangfeng},
	month = jul,
	year = {2011},
	pages = {020501},
}

@article{kassal_polynomial-time_2008,
	title = {Polynomial-time quantum algorithm for the simulation of chemical dynamics},
	volume = {105},
	url = {https://www.pnas.org/doi/abs/10.1073/pnas.0808245105},
	doi = {10.1073/pnas.0808245105},
	number = {48},
	urldate = {2024-07-16},
	journal = {Proceedings of the National Academy of Sciences},
	author = {Kassal, Ivan and Jordan, Stephen P. and Love, Peter J. and Mohseni, Masoud and Aspuru-Guzik, Alán},
	month = dec,
	year = {2008},
	pages = {18681--18686},
}

@article{pruisken_universal_1988,
	title = {Universal {Singularities} in the {Integral} {Quantum} {Hall} {Effect}},
	volume = {61},
	url = {https://link.aps.org/doi/10.1103/PhysRevLett.61.1297},
	doi = {10.1103/PhysRevLett.61.1297},
	number = {11},
	urldate = {2024-08-02},
	journal = {Physical Review Letters},
	author = {Pruisken, A. M. M.},
	month = sep,
	year = {1988},
	pages = {1297--1300},
}

@article{daley_practical_2022,
	title = {Practical quantum advantage in quantum simulation},
	volume = {607},
	rights = {2022 Springer Nature Limited},
	issn = {1476-4687},
	url = {https://www.nature.com/articles/s41586-022-04940-6},
	doi = {10.1038/s41586-022-04940-6},
	pages = {667--676},
	number = {7920},
	journal = {Nature},
	author = {Daley, Andrew J. and Bloch, Immanuel and Kokail, Christian and Flannigan, Stuart and Pearson, Natalie and Troyer, Matthias and Zoller, Peter},
	urldate = {2026-09-04},
	year = {2022},
    month = {07},
	langid = {english},
}

@article{seye_localization_2025,
	title = {Localization structure of electronic states in the quantum Hall effect},
	volume = {112},
	url = {https://link.aps.org/doi/10.1103/266g-9zc3},
	doi = {10.1103/266g-9zc3},
	pages = {144202},
	number = {14},
	journal = {Physical Review B},
	shortjournal = {Phys. Rev. B},
	author = {Seye, Alioune and Filoche, Marcel},
	urldate = {2026-09-04},
	year = {2025},
    month = {10},
    day = {07},
}

@article{joynt_conditions_1984,
	title = {Conditions for the quantum Hall effect},
	volume = {29},
	url = {https://link.aps.org/doi/10.1103/PhysRevB.29.3303},
	doi = {10.1103/PhysRevB.29.3303},
	pages = {3303--3317},
	number = {6},
	journal = {Physical Review B},
	shortjournal = {Phys. Rev. B},
	author = {Joynt, Robert and Prange, R. E.},
	urldate = {2026-09-04},
	year = {1984},
    month = {03},
    day = {15},
}

@article{ciardi_finite-temperature_2022,
	title = {Finite-temperature phases of trapped bosons in a two-dimensional quasiperiodic potential},
	volume = {105},
	url = {https://link.aps.org/doi/10.1103/PhysRevA.105.L011301},
	doi = {10.1103/PhysRevA.105.L011301},
	number = {1},
	urldate = {2024-09-25},
	journal = {Physical Review A},
	author = {Ciardi, Matteo and Macr{\`i}, Tommaso and Cinti, Fabio},
	month = jan,
	year = {2022},
	pages = {L011301},
}

@article{zhu_thermodynamic_2023,
	title = {Thermodynamic {Phase} {Diagram} of {Two}-{Dimensional} {Bosons} in a {Quasicrystal} {Potential}},
	volume = {130},
	url = {https://link.aps.org/doi/10.1103/PhysRevLett.130.220402},
	doi = {10.1103/PhysRevLett.130.220402},
	number = {22},
	urldate = {2024-09-25},
	journal = {Physical Review Letters},
	author = {Zhu, Zhaoxuan and Yao, Hepeng and Sanchez-Palencia, Laurent},
	month = may,
	year = {2023},
	pages = {220402},
}

@article{bertoli_finite-temperature_2018,
	title = {Finite-{Temperature} {Disordered} {Bosons} in {Two} {Dimensions}},
	volume = {121},
	url = {https://link.aps.org/doi/10.1103/PhysRevLett.121.030403},
	doi = {10.1103/PhysRevLett.121.030403},
	number = {3},
	urldate = {2024-09-25},
	journal = {Physical Review Letters},
	author = {Bertoli, G. and Michal, V. P. and Altshuler, B. L. and Shlyapnikov, G. V.},
	month = jul,
	year = {2018},
	pages = {030403},
}

@article{michal_finite-temperature_2016,
	title = {Finite-temperature fluid–insulator transition of strongly interacting {1D} disordered bosons},
	volume = {113},
	url = {https://www.pnas.org/doi/abs/10.1073/pnas.1606908113},
	doi = {10.1073/pnas.1606908113},
	number = {31},
	urldate = {2024-09-25},
	journal = {Proceedings of the National Academy of Sciences},
	author = {Michal, Vincent P. and Aleiner, Igor L. and Altshuler, Boris L. and Shlyapnikov, Georgy V.},
	month = aug,
	year = {2016},
	pages = {E4455--E4459},
}

@article{man_experimental_2005,
	title = {Experimental measurement of the photonic properties of icosahedral quasicrystals},
	volume = {436},
	copyright = {2005 Springer Nature Limited},
	issn = {1476-4687},
	url = {https://www.nature.com/articles/nature03977},
	doi = {10.1038/nature03977},
	number = {7053},
	urldate = {2024-09-25},
	journal = {Nature},
	author = {Man, Weining and Megens, Mischa and Steinhardt, Paul J. and Chaikin, P. M.},
	month = aug,
	year = {2005},
	pages = {993--996},
}

@article{spurrier_semiclassical_2018,
	title = {Semiclassical dynamics, {Berry} curvature, and spiral holonomy in optical quasicrystals},
	volume = {97},
	url = {https://link.aps.org/doi/10.1103/PhysRevA.97.043603},
	doi = {10.1103/PhysRevA.97.043603},
	number = {4},
	urldate = {2024-10-01},
	journal = {Physical Review A},
	author = {Spurrier, Stephen and Cooper, Nigel R.},
	month = apr,
	year = {2018},
	pages = {043603},
}

@article{gerbier_expansion_2008,
  title = {Expansion of a Quantum Gas Released from an Optical Lattice},
  author = {Gerbier, F. and Trotzky, S. and F\"olling, S. and Schnorrberger, U. and Thompson, J. D. and Widera, A. and Bloch, I. and Pollet, L. and Troyer, M. and Capogrosso-Sansone, B. and Prokof'ev, N. V. and Svistunov, B. V.},
  journal = {Phys. Rev. Lett.},
  volume = {101},
  issue = {15},
  pages = {155303},
  numpages = {4},
  year = {2008},
  month = {Oct},
  publisher = {American Physical Society},
  doi = {10.1103/PhysRevLett.101.155303},
  url = {https://link.aps.org/doi/10.1103/PhysRevLett.101.155303}
}

@article{martirosyan_universal_2025,
	title = {A universal speed limit for spreading of coherence},
	volume = {647},
	rights = {2025 The Author(s)},
	issn = {1476-4687},
	url = {https://www.nature.com/articles/s41586-025-09735-z},
	doi = {10.1038/s41586-025-09735-z},
	pages = {608--612},
	number = {8090},
	journal = {Nature},
	author = {Martirosyan, Gevorg and Gazo, Martin and Etrych, Jiří and Fischer, Simon M. and Morris, Sebastian J. and Ho, Christopher J. and Eigen, Christoph and Hadzibabic, Zoran},
	urldate = {2025-11-21},
	year = {2025},
    month = {11},
	langid = {english},
}

@article{manovitz_quantum_2025,
	title = {Quantum coarsening and collective dynamics on a programmable simulator},
	volume = {638},
	rights = {2025 The Author(s)},
	issn = {1476-4687},
	url = {https://www.nature.com/articles/s41586-024-08353-5},
	doi = {10.1038/s41586-024-08353-5},
	pages = {86--92},
	number = {8049},
	journal = {Nature},
	author = {Manovitz, Tom and Li, Sophie H. and Ebadi, Sepehr and Samajdar, Rhine and Geim, Alexandra A. and Evered, Simon J. and Bluvstein, Dolev and Zhou, Hengyun and Koyluoglu, Nazli Ugur and Feldmeier, Johannes and Dolgirev, Pavel E. and Maskara, Nishad and Kalinowski, Marcin and Sachdev, Subir and Huse, David A. and Greiner, Markus and Vuletić, Vladan and Lukin, Mikhail D.},
	urldate = {2025-11-21},
	year = {2025},
    month = {02},
	langid = {english},
}

@article{navon_emergence_2016,
	title = {Emergence of a turbulent cascade in a quantum gas},
	volume = {539},
	rights = {2016 Macmillan Publishers Limited, part of Springer Nature. All rights reserved.},
	issn = {1476-4687},
	url = {https://www.nature.com/articles/nature20114},
	doi = {10.1038/nature20114},
	pages = {72--75},
	number = {7627},
	journal = {Nature},
	author = {Navon, Nir and Gaunt, Alexander L. and Smith, Robert P. and Hadzibabic, Zoran},
	urldate = {2025-11-21},
	year = {2016},
    month ={11},
	langid = {english},
}

@article{etrych_pinpointing_2023,
	title = {Pinpointing Feshbach resonances and testing Efimov universalities in {K} 39},
	volume = {5},
	issn = {2643-1564},
	url = {https://link.aps.org/doi/10.1103/PhysRevResearch.5.013174},
	doi = {10.1103/PhysRevResearch.5.013174},
	pages = {013174},
	number = {1},
	journal = {Physical Review Research},
	shortjournal = {Phys. Rev. Research},
	author = {Etrych, Jiří and Martirosyan, Gevorg and Cao, Alec and Glidden, Jake A. P. and Dogra, Lena H. and Hutson, Jeremy M. and Hadzibabic, Zoran and Eigen, Christoph},
	urldate = {2024-10-24},
	year = {2023},
    month = {03},
    day = {13},
	langid = {english},
}

@article{langen_ultracold_2015,
	title = {Ultracold Atoms Out of Equilibrium},
	volume = {6},
	issn = {1947-5462},
	url = {https://www.annualreviews.org/content/journals/10.1146/annurev-conmatphys-031214-014548},
	doi = {10.1146/annurev-conmatphys-031214-014548},
	pages = {201--217},
	journal = {Annual Review of Condensed Matter Physics},
	author = {Langen, Tim and Geiger, Remi and Schmiedmayer, Jörg},
	urldate = {2026-08-10},
	year = {2015},
}

@article{gottlob_schneider_2023,
    title = {Hubbard Hamiltonian for the 8fold Optical Quasicrystal},
    volume = {},
    url = {https://www.repository.cam.ac.uk/handle/1810/348940},
    doi = {10.17863/CAM.95664},
    pages = {},
    journal = {10.17863/CAM.95664},
    publisher = {Apollo - University of Cambridge Repository},
    author = {Gottlob, Emmanuel and Schneider, Ulrich},
    year = {2023}
}

@book{walter_crystallography_2009,
	location = {Heidelberg},
	edition = {1},
	title = {Crystallography of Quasicrystals},
	isbn = {978-3-642-01898-5},
	url = {https://link.springer.com/book/10.1007/978-3-642-01899-2},
	series = {Springer Series in Materials Science},
	publisher = {Springer Berlin},
	author = {Walter, Steurer and Deloudi, Sofia},
	year = {2009},
    month={09},
    day={11},
	doi = {10.1007/978-3-642-01899-2},
}

@article{schafer_tools_2020,
	title = {Tools for quantum simulation with ultracold atoms in optical lattices},
	volume = {2},
	rights = {2020 Springer Nature Limited},
	issn = {2522-5820},
	url = {https://www.nature.com/articles/s42254-020-0195-3},
	doi = {10.1038/s42254-020-0195-3},
	pages = {411--425},
	number = {8},
	journal = {Nature Reviews Physics},
	shortjournal = {Nat Rev Phys},
	author = {Schäfer, Florian and Fukuhara, Takeshi and Sugawa, Seiji and Takasu, Yosuke and Takahashi, Yoshiro},
	urldate = {2026-08-14},
	year = {2020},
    month={08},
	langid = {english},
}

@article{
doi:10.1073/pnas.1408861112,
author = {Simon Braun  and Mathis Friesdorf  and Sean S. Hodgman  and Michael Schreiber  and Jens Philipp Ronzheimer  and Arnau Riera  and Marco del Rey  and Immanuel Bloch  and Jens Eisert  and Ulrich Schneider },
title = {Emergence of coherence and the dynamics of quantum phase transitions},
journal = {Proceedings of the National Academy of Sciences},
volume = {112},
number = {12},
pages = {3641-3646},
year = {2015},
doi = {10.1073/pnas.1408861112},
URL = {https://www.pnas.org/doi/abs/10.1073/pnas.1408861112}}

@article{murthy_observation_2015,
	title = {Observation of the Berezinskii-Kosterlitz-Thouless Phase Transition in an Ultracold Fermi Gas},
	volume = {115},
	doi = {10.1103/PhysRevLett.115.010401},
	number = {1},
	journal = {Physical Review Letters},
	shortjournal = {Phys. Rev. Lett.},
	author = {Murthy, P. A.},
	year = {2015},
}

@article{murthy_quantum_2019,
	title = {Quantum scale anomaly and spatial coherence in a 2D Fermi superfluid},
	volume = {365},
	url = {https://www.science.org/doi/10.1126/science.aau4402},
	doi = {10.1126/science.aau4402},
	pages = {268--272},
	number = {6450},
	journal = {Science},
	author = {Murthy, Puneet A. and Defenu, Nicolò and Bayha, Luca and Holten, Marvin and Preiss, Philipp M. and Enss, Tilman and Jochim, Selim},
	urldate = {2026-07-14},
	year = {2019},
    month={07},
    day={19},
}

@article{boettcher_quasi-long-range_2016,
	title = {Quasi-long-range order in trapped two-dimensional Bose gases},
	volume = {94},
	issn = {2469-9926, 2469-9934},
	url = {http://arxiv.org/abs/1605.00597},
	doi = {10.1103/PhysRevA.94.011602},
	pages = {011602},
	number = {1},
	journal = {Physical Review A},
	shortjournal = {Phys. Rev. A},
	author = {Boettcher, Igor and Holzmann, Markus},
	urldate = {2026-07-14},
	year = {2016},
    month={07},
    day={11},
	eprinttype = {arxiv},
	eprint = {1605.00597 [cond-mat.quant-gas]},
}

@article{guo_observation_2024,
	title = {Observation of the 2D–1D crossover in strongly interacting ultracold bosons},
	volume = {20},
	rights = {2024 The Author(s), under exclusive licence to Springer Nature Limited},
	issn = {1745-2481},
	url = {https://www.nature.com/articles/s41567-024-02459-3},
	doi = {10.1038/s41567-024-02459-3},
	pages = {934--938},
	number = {6},
	journal = {Nature Physics},
	shortjournal = {Nat. Phys.},
	author = {Guo, Yanliang and Yao, Hepeng and Ramanjanappa, Satwik and Dhar, Sudipta and Horvath, Milena and Pizzino, Lorenzo and Giamarchi, Thierry and Landini, Manuele and Nägerl, Hanns-Christoph},
	urldate = {2026-07-14},
	year = {2024},
    month={06},
	langid = {english},
}

@article{song_realizing_2022,
	title = {Realizing discontinuous quantum phase transitions in a strongly correlated driven optical lattice},
	volume = {18},
	rights = {2022 The Author(s), under exclusive licence to Springer Nature Limited},
	issn = {1745-2481},
	url = {https://www.nature.com/articles/s41567-021-01476-w},
	doi = {10.1038/s41567-021-01476-w},
	pages = {259--264},
	number = {3},
	journal = {Nature Physics},
	shortjournal = {Nat. Phys.},
	author = {Song, Bo and Dutta, Shovan and Bhave, Shaurya and Yu, Jr-Chiun and Carter, Edward and Cooper, Nigel and Schneider, Ulrich},
	urldate = {2024-02-21},
	year = {2022},
    month={03},
	langid = {english},
}
\end{document}